\documentclass{article}

\usepackage{arxiv}
\usepackage[utf8]{inputenc} 
\usepackage[T1]{fontenc}    
\usepackage[colorlinks=true, allcolors=blue]{hyperref}
\usepackage{booktabs}       
\usepackage{amsfonts}       
\usepackage{nicefrac}       
\usepackage{microtype}      
\usepackage[version=4]{mhchem}
\usepackage{amsmath}
\usepackage{makecell}       
\usepackage{graphicx}       
\usepackage{subcaption}     
\usepackage{float}          
\usepackage[colorinlistoftodos]{todonotes}
\usepackage{listings}
\usepackage{xcolor}
\usepackage{makecell}       
\usepackage{comment}
\usepackage[normalem]{ulem} 
\usepackage{colortbl}       

\hypersetup{
   breaklinks=true,   
}

\definecolor{LightGreen}{HTML}{95F985}
\definecolor{DarkGreen}{HTML}{00AB08}
\definecolor{codegreen}{rgb}{0,0.6,0}
\definecolor{codegray}{rgb}{0.5,0.5,0.5}
\definecolor{codepurple}{rgb}{0.58,0,0.82}
\definecolor{backcolour}{rgb}{0.95,0.95,0.92}

\lstdefinestyle{mystyle}{
    backgroundcolor=\color{backcolour},   
    commentstyle=\color{codegreen},
    keywordstyle=\color{magenta},
    numberstyle=\tiny\color{codegray},
    stringstyle=\color{codepurple},
    basicstyle=\ttfamily\footnotesize,
    breakatwhitespace=false,         
    breaklines=true,                 
    captionpos=b,                    
    keepspaces=true,                 
    numbers=left,                    
    numbersep=5pt,                  
    showspaces=false,                
    showstringspaces=false,
    showtabs=false,                  
    tabsize=2
}

\title{Robust and scalable uncertainty estimation with conformal prediction for machine-learned interatomic potentials}

\author{
  Yuge Hu\textsuperscript{1}, Joseph Musielewicz\textsuperscript{2}, Zachary Ulissi\textsuperscript{2}, Andrew J. Medford\textsuperscript{1, $\dagger$}\\
  \textsuperscript{1} Department of Chemical and Biomolecular Engineering\\
  Georgia Institute of Technology\\
  \textsuperscript{2} Department of Chemical Engineering\\
  Carnegie Mellon University\\
  \textsuperscript{$\dagger$} \textit{Corresponding author: Andrew J. Medford, ajm@gatech.edu}  \\
}

\begin{document}
\maketitle

\begin{abstract}
\textbf{Abstract}

Uncertainty quantification (UQ) is important to machine learning (ML) force fields to assess the level of confidence during prediction, as ML models are not inherently physical and can therefore yield catastrophically incorrect predictions. Established \textit{a-posteriori} UQ methods, including ensemble methods, the dropout method, the delta method, and various heuristic distance metrics, have limitations such as being computationally challenging for large models due to model re-training. In addition, the uncertainty estimates are often not rigorously calibrated. In this work, we propose combining the distribution-free UQ method, known as conformal prediction (CP), with the distances in the neural network's latent space to estimate the uncertainty of energies predicted by neural network force fields. We evaluate this method (CP+latent) along with other UQ methods on two essential aspects, calibration, and sharpness, and find this method to be both calibrated and sharp under the assumption of independent and identically-distributed (i.i.d.) data. We show that the method is relatively insensitive to hyperparameters selected, and test the limitations of the method when the i.i.d. assumption is violated. Finally, we demonstrate that this method can be readily applied to trained neural network force fields with traditional and graph neural network architectures to obtain estimates of uncertainty with low computational costs on a training dataset of 1 million images to showcase its scalability and portability. Incorporating the CP method with latent distances offers a calibrated, sharp and efficient strategy to estimate the uncertainty of neural network force fields. In addition, the CP approach can also function as a promising strategy for calibrating uncertainty estimated by other approaches.

\end{abstract}

\keywords{machine learning force fields \and neural network force fields \and uncertainty quantification \and conformal prediction}

\section{Introduction}

Machine learning (ML) models have been applied extensively to approximate the potential energy surfaces that govern atomistic interactions. These models, known as ML force fields (MLFFs) or ML interatomic potentials, have been widely applied in molecular simulations of elemental systems\cite{Liu2020SingleNN:Transferability,Bartok2013OnEnvironments,Behler2007GeneralizedSurfacesc,Gao2018ModelingDynamics,Boes2017NeuralSurface}, organic molecules\cite{Smith2017ANI-1:Cost,Chmiela2017MachineFields,Schutt2018SchNet-AMaterials} metal oxides\cite{Artrith2011High-dimensionalOxide,Wang2018AZirconia}, bulk solvent molecules\cite{Singraber2019Library-BasedPotentials}, solid-solvent interfaces\cite{Natarajan2016NeuralSurfaces} and catalytic systems\cite{Khorshidi2016Amp:Simulations,Chanussot2021OpenChallenges,Shuaibi2021RotationConvolutions,Musielewicz2022FINETUNA:Simulations}. These MLFFs typically use quantum mechanical (QM) calculations as training data, such as density functional theory (DFT)\cite{Liu2020SingleNN:Transferability,Bartok2013OnEnvironments,Smith2017ANI-1:Cost,Chanussot2021OpenChallenges,Behler2021MachinePerspective}, symmetry adapted perturbation theory\cite{Metcalf2020ApproachesTheory,Schriber2021CLIFF:Field}, CCSD-T\cite{Smith2019ApproachingLearning,Schran2020AutomatedGround}, and quantum Monte Carlo\cite{Tirelli2021HighCarlo}. ML force fields provide a promising strategy to balance the speed and accuracy of predictions, and exhibit good generality in predicting a wide range of chemical phenomena such as bond breaking and multiple types of bonding. ML force fields allow researchers to interpolate with high fidelity, bypassing the major computational bottleneck of solving the electronic structures repeatedly. ML force fields have the advantage of high accuracy compared to conventional empirical force field methods yet at significantly lower computational costs than QM calculations\cite{Behler2021MachinePerspective}. Typical ML force fields include non-parametric kernel methods, such as sGDML\cite{Chmiela2019SGDML:Learning} and GAP-SOAP\cite{Bartok2013OnEnvironments,Bartok2015GaussianIntroduction}, parametric neural network models with fixed features (e.g. ANI\cite{Smith2017ANI-1:Cost}, BPNN\cite{Behler2015ConstructingReview}, DPMD\cite{Zhang2018DeepMechanics}, SingleNN\cite{Liu2020SingleNN:Transferability}, GMP\cite{Lei2021ASystems}) and deep learning graph-based methods with learned features (e.g. PhysNet\cite{Unke2019PhysNet:Charges}, SchNet\cite{Schutt2018SchNet-AMaterials}, DimeNet++\cite{Gasteiger2020FastMolecules}, SpinConv Net\cite{Shuaibi2021RotationConvolutions}, GemNet-T\cite{Gasteiger2021GemNet:Molecules})\cite{Pinheiro2021ChoosingPotential}.


 Unlike empirical force fields that are built on explicit mathematical expressions, most MLFFs have few if any physical constraints, such as short-range repulsion or zero interaction at long range. Instead, MLFFs rely on the model's flexibility to faithfully interpolate the space defined by the training data. When running into observations that are poorly represented in the training data, the accuracy of the ML models can no longer be guaranteed.
Uncertainty quantification (UQ) establishes the confidence associated with the model's prediction, and knowing the uncertainty is vital to interpreting the model's predictions\cite{Tran2020MethodsPredictions, Peterson2017AddressingLearning,Pernot2022TheChemistry}. 
Additionally, the high dimensionality of the space defined by the chemical systems of interest can be large (10's - 1000's of dimensions), making it impractical to densely sample the potential energy surface. Integrating UQ into active learning schemes allows robust re-training via identification of datapoints with the highest uncertainty, allowing for a balance of exploration and exploitation\cite{Tran2020MethodsPredictions,Tran2018DynamicScience,Smith2021AutomatedAluminum,Peterson2017AddressingLearning,Pernot2022TheChemistry}. However, the utility of any UQ scheme is ultimately determined by its reliability, since inaccurate uncertainty estimates may lead to incorrect interpretations of a model or inefficient and inaccurate active learning models.

To asses uncertainty quantification approaches, it is first necessary to consider the underlying sources of errors. The errors in chemical simulation and computation are categorized into three types: numerical errors, model errors, and parametric errors\cite{Wen2020UncertaintyPotentials, Pernot2022TheChemistry}. Numerical errors refer to the errors generated by using finite arithmetic implementations or the random errors incurred by stochastic processes within computational algorithms. Model errors are associated with inaccuracies due to the level of theory or the choice of basis set. Parametric errors arise from the statistical estimation of the model parameters with respect to the training data. Numerical errors are usually considered well-controlled in computational chemistry and therefore neglected, while model errors are unavoidable but reasonably well understood for a given level of theory. Thus, this work is focused on UQ for addressing the parametric errors that are associated with the neural network (or other ML models) models as a result of statistical estimation of parameters during the training process. In addition to the magnitude of these errors, it is also worth considering their distribution. Pernot points out that parametric errors depend on the functional form of regression models and uncertainty metrics, and therefore are not always Gaussian\cite{Pernot2022TheChemistry}, making it necessary to avoid distributional assumptions to allow for robust UQ on these errors. 


Numerous UQ methods have been developed in the fields of statistics, ML, and computational chemistry. The approaches can be broadly classified as model-specific methods, where the UQ estimate is part of the model structure, and more generic methods that can be applied to different model types. Model-specific methods include uncertainty estimation of Gaussian processes and Bayesian neural networks, whose explicit mathematical expressions of uncertainty are readily available. Another example is dropout neural networks \cite{Zhan2021UncertaintyModels,Nilsen2022EpistemicMethod,Endo2015ConfidenceMethod,Wen2020UncertaintyPotentials,MoriartyUnlockNN:Systems} that provide uncertainty estimates by computing the standard deviation of a collection of models, each with different nodes dropped out during training and prediction. Examples of generic methods include ensemble methods that create an ensemble of models and use the spread of the outcomes as the estimate of uncertainty \cite{Peterson2017AddressingLearning,Tran2020MethodsPredictions,Smith2021AutomatedAluminum}, the delta method to estimate the approximate standard error \cite{Zhan2021UncertaintyModels,Nilsen2022EpistemicMethod,Du2021UncertaintyNetworks,Palmer2021CalibratedModels} and \textit{a posteriori} distance-based methods \cite{Pernot2022TheChemistry,PaulJanet2019} that use distance to training points as a heuristic for uncertainty. The \textit{a-posteriori} distance-based methods are built upon the idea that the errors are correlated with the distances between new observations (test data points) and the training data --- shorter distance corresponds to more similarity, therefore less uncertainty, and \textit{vice versa}\cite{PaulJanet2019,Botu2017MachineOutlook}. This type of \textit{a-posteriori} approach requires a separate set of calibration data to correlate the distance heuristics with the errors, typically based on the assumption that errors are normally distributed \cite{PaulJanet2019}.

The decision of which UQ method to adopt depends on the application and the complexity of the employed models and datasets, as there is no consensus on a general best approach \cite{Peterson2017AddressingLearning,Zhan2021UncertaintyModels,Wen2020UncertaintyPotentials,Tran2020MethodsPredictions}. Model-specific UQ approaches require a specific form of the ML model and are therefore less flexible, and generally require model re-training, so in this work we focus on the more generic approaches. The ensemble method is one of the most popular and intuitive approaches for estimating uncertainty of MLFFs, and has already been adopted in multiple instances of active learning schemes \cite{Musielewicz2022FINETUNA:Simulations,Gasteiger2020FastMolecules}. One major drawback of the ensemble method is that the computational costs increase with the model complexity due to re-training, making it cost-prohibitive when applied to complex models and large datasets, although techniques such as implicit ensembles based on dropout may be used to mitigate this issue \cite{Ganaie2022EnsembleReview}. Additionally, the ensemble variances on configurations that are distinct from the training data are likely to approach zero, potentially leading to unreliable uncertainty estimates for systems that are very different from the training data \cite{Liu2019MolecularCompounds}. The dropout method also has shortcomings, such as varying levels of model performance depending on the choice of dropout rate and requiring additional effort to optimize hyperparameters and evaluate the resulting models. The delta method also has its limit for very large models as it requires computing the Hessian of the loss function with respect to model parameters and inverting the resulting Hessian matrix, both of which require significant computational effort in the limit of many model inputs or complex model architectures. Moreover, the delta method utilizes assumptions about model linearity and assumes normally distributed errors \cite{Zhan2021UncertaintyModels}. Distance-based methods are more scalable and can be applied readily to pre-trained models, and tend to have better performance with out-of-domain samples compared to the ensemble method \cite{Liu2019MolecularCompounds}. However, one inherent disadvantage is that distance-based heuristics are not in the same unit as errors, thus requiring assumptions about the form of the error distribution (which typically assumed to be Gaussian) \cite{PaulJanet2019}. The reliance on the assumption of normally distributed errors is a common weakness of many UQ techniques, since the error types encountered in computational chemistry are generally not normal\cite{Pernot2022TheChemistry} as discussed above. The violation of this underlying assumption may lead to systemic under- and over-confidence when applying these UQ methods. Indeed, Tran et. al. report that various UQ methods considered, such as the ensemble and sampling dropout methods, are not inherently calibrated and introduce a re-calibration scheme as a remedy \cite{Tran2020MethodsPredictions}. While re-calibration improves the reliability of the UQ estimates, it also creates additional conceptual complexity and increases the computational cost.

In this work, we propose a UQ method that integrates the distribution-free statistical framework, conformal prediction (CP), with the previously published \textit{a-posteriori} latent-distance-based method \cite{PaulJanet2019} to relax the assumption of normally distributed errors. While the statistical rigor of CP guarantees calibration, this method also fully leverages the advantage of the distance-based uncertainty heuristic to be applied to complex pre-trained models, as the cost of uncertainty estimation does not scale with the model complexity. Together, the CP method integrated with the \textit{a-posteriori} latent-distance metric offers a reliable and scalable pathway to estimate the uncertainty for a trained neural network model, and can easily be adapted to other types of machine-learning models by utilizing distances in feature space. We show that this approach is well-calibrated regardless of the nature of the error distribution or choice of hyperparameters, has a computational cost that is much smaller than evaluating the neural network model, and can be applied to advanced deep learning models such as graph neural networks that have already been trained on massive datasets. 

\section{Methods}

\subsection{Reference data}
\label{sec:reference_data}

We conduct the analysis by applying CP to neural network force fields built on three benchmark datasets, MD17-Aspirin\cite{Chmiela2017MachineFields}, QM9\cite{Ramakrishnan2014QuantumMolecules}, and the Open Catalyst Project (OC20)\cite{Chanussot2021OpenChallenges}, with an increasing number of training data and chemical complexity (number of elements, types of bonding). We focus the comparison across several different UQ methods on MD17-Aspirin and QM9 since it is impractical to train most models on the large OC20 set. The MD17-Aspirin dataset consists of 211K images in the AIMD trajectory of an aspirin molecule computed with the PBE+vdW-TS electronic level of theory\cite{Chmiela2017MachineFields}. The QM9 dataset has 130K small organic chemical molecules made up of C, H, O, N, F, generated at the B3LYP/6-31G(2df,p) level of theory\cite{Ramakrishnan2014QuantumMolecules}. In Section \ref{sec:ood} we further break the QM9 dataset into a dataset with all molecules with fluorine and another dataset made up of only C, H, O, and N to test the out-of-distribution scenario. Finally, we test the scalability of the CP method on neural network force field models trained with the OC20 S2EF dataset made up of 55 elements and 82 adsorbates\cite{Chanussot2021OpenChallenges}.

\subsection{Gaussian multi-pole (GMP) featurization and SingleNN neural network architecture}
\label{sec:GMPs}  

 ML models for all benchmark datasets are constructed using feed-forward neural networks \cite{Behler2015ConstructingReview} with fixed local descriptors. We utilize the GMP featurization scheme as described by Lei and Medford\cite{Lei2021ASystems} as the atomistic fingerprinting scheme. GMP uses multipole expansions to describe the reconstructed electronic density around every central atom and its neighbors to encode local environments. There are two main hyperparameters in GMP featurization scheme: a vector of ``radial probe'' distances ($\sigma$) and the maximum order of the multipole expansion. The vector of radial probe distances defines the resolution in the radial coordinate, while the maximum multipole order defines resolution in the angular coordinate. GMP descriptors use a set of fitted primitive Gaussian functions to encode the information about chemical species and, therefore, naturally allow interpolation among element types. Instead of separate neural network models for every element type, we adopted the SingleNN (SNN) architecture of Liu and Kitchin\cite{Liu2020SingleNN:Transferability},  which shares a neural network latent space across all element types in the training data. The combination of GMP descriptors and SingleNN models allows the models to make predictions on unseen elements, a feature which we use to study the out-of-distribution samples for UQ analysis in Sec. \ref{sec:ood} on QM9 dataset. In the GMP+SNN model architecture each atom has a latent (or feature) vector, and these vectors are averaged over each system to provide a single vector per system that is used as input to the CP model. We note that the use of averaged fingerprints is related to a limitation of the CP method, since it can only predict uncertainties at the system level due to the lack of a ground truth for energy per atom. Due to the large size of OC20 and practical computational limitations, the pre-trained GMP+SNN model is trained on 1M adsorbate-catalyst systems uniformly sampled at random from the original 20M S2EF dataset.

\subsection{Graph convolution neural network GemNet-OC}

For the OC20 dataset, we extend the application of the proposed CP+latent method to a graph neural network (GNN) whose interaction blocks are used as latent representations.
GNNs have been designed to learn generalisable representations of atomic structures using graphical representations. 
The specific GNN we use in this work is GemNet-OC, pre-trained on the entire 134M OC20 dataset, which encodes atoms as nodes in a graph, and the interactions between them as edges. 
When trained on the OC20 dataset GemNet-OC has been shown to be among the most effective general ML interatomic potentials for predicting energies and forces, according to the OC20 leaderboard \cite{Chanussot2021OpenChallenges}.
GemNet-OC is useful for its latent representation because of the way it encodes information.
It encodes distances, angular, and dihedral information using Bessel basis functions. 
For energy predictions, this representation is invariant with respect to global rotations, while preserving relative rotational information. 
In a series of interactions blocks, GemNet-OC learns to pass geometric information as messages between edges.
This results in a valuable latent representation of atomic structures, which is then transformed into an energy contribution by each of the model's output blocks \cite{Gasteiger2021GemNet:Molecules,GemNet-OC:Datasets}.
This latent representation should preserve similarities between structures in a way that is conducive to UQ using the CP method.
The GemNet-OC latent representation takes the form of an $M \times N$ dimensional matrix, where $M$ is the number of atoms in the system (and the number of nodes in the graph) and $N$ is the width of the interaction block (which is arbitrary, but affects the expressiveness of the model).
We averaged this matrix to a vector of image-wise representation over the number of atoms for the purposes of distance computation in the CP method.

\subsection{Metrics for calibration and sharpness}

To characterize and quantify the performance of a UQ method, we introduce two important concepts: \textit{calibration} and \textit{sharpness}\cite{Pernot2022TheChemistry, Tran2018ActiveEvolution, Kuleshov2018AccurateRegression}. Following the convention from Pernot\cite{Pernot2022TheChemistry}, \textit{calibration} is defined as whether the confidence, either based on distributional assumptions or statistical analysis, is true to the actual test probability of containing the ground truth. For example, the prediction sets with 95\% confidence that contain the ground truth 95\% of the time are calibrated. Calibration guarantees the reliability of the UQ method, but calibration alone is not sufficient to establish a valid and useful UQ method. Another important factor regarding the usefulness of the UQ method is sharpness. \textit{Sharpness} is defined as the ``tightness'' of the prediction sets (i.e. the size of the error bars on predictions). While there is no absolute reference value for sharpness, the sense of sharpness is usually established by comparing different methods under the same confidence. A calibrated UQ method with narrower prediction bandwidths (i.e. predicted error bars) is generally preferred over a method with broader prediction bandwidths. We note that sharpness is conditioned on calibration: a fair comparison for sharpness metrics cannot be drawn unless the models are well-calibrated and under the same confidence level. 

To test for calibration, we adopt two methods. One method is the visual comparison of the prediction sets with different uncertainty metrics at the expected confidence levels of 68\% and 95\%, equivalent to $\approx$ 1 and 2 $\sigma$'s if errors are normally distributed. Another method is the calibration plot as defined in Kuleshov et al.\cite{Kuleshov2018AccurateRegression, Tran2020MethodsPredictions} 
For sharpness ($sha$), we follow the definition as published by Tran et al.\cite{Tran2020MethodsPredictions}:

\begin{equation*}
    sha = \sqrt{\frac{1}{N} \sum_{n=1}^{N} var(F_{n})}
\end{equation*}

where $var(F_{n})$ is the variance of random variables $n$ whose cumulative distribution function is $F_{n}$. Taking the square root makes $sha$ have the same unit as the quantity of interest (energy in this case). The definition indicates that sharpness is linearly correlated with the average of standard deviations of predictions. We denote $sha_{\sigma}$ for statistical methods that calculate the standard deviation $\sigma$, while for CP which produces prediction sets given an expected confidence level $x\%$, we denote $sha_{x\%}$. For example, $sha_{68\%}$ is approximately equal to $sha_{\sigma}$ in the case of a normal distribution. By this definition of sharpness, we would prefer a calibrated method with a smaller numerical $sha$ value.

\subsection{UQ methods for comparison}

Three common alternative UQ methods are chosen for comparison: the ensemble method, the dropout method, and the distance-based method with the normal distribution assumption. We utilize pragmatic choices in constructing the ensemble and dropout methods with an emphasis on efficiency (small ensembles). We note that the results of these techniques may be greatly improved with larger ensembles and/or tuning of hyperparameters, but counter this by showing that the results of the CP method are largely independent of hyperparameter choices. Thus, we expect that the results provide a baseline for the expected performance of these techniques in the limit where limited effort is devoted to tuning hyperparamters of the UQ model and the computational cost of UQ is comparable to the cost of model evaluation.

\textbf{Ensemble Method}

We used an ensemble of 4 different standard feed-forward neural network architectures on the same set of training data to calculate the standard deviation of errors as an uncertainty estimate. The 4 different neural network architectures are [64,64,64] (baseline), [128,128,128], [128,64,32,16,8], and [64,32,16,8] for MD17-Aspirin and QM9 datasets, where the notation [$a$, $b$, ..., $n$] corresponds to a neural network with a total number of layers equal to the length of the list, and $a$ nodes in the first layer, $b$ nodes in the second layer, and $n$ nodes in the final layer. The NN architecture [64,64,64] is chosen as the baseline because it is closest to the architecture that yielded the best results for energy and force training ([50, 50, 50]) as published in a previous publication by Lei and Medford on the GMP descriptors\cite{Lei2021ASystems} for MD17-Aspirin dataset. Other architectures are selected arbitrarily as reasonable perturbations of this structure.

\textbf{Dropout Method}

We used an ensemble of 4 dropout neural networks (\texttt{dropout\_rate} = 0.2), each with different initial weight randomization, on the same set of training data to calculate the standard deviation of errors as an uncertainty estimate. The dropout neural networks all have the same architecture as the baseline model --- [64,64,64]. 

\textbf{Negative log-likelihood Method with heuristic distance metrics}

Janet et. al. introduced the Negative log-likelihood (NLL) method as a calibration approach to estimate the predicted variance for a calculated distance heuristic $d$ based on the assumption that the errors are normally distributed\cite{PaulJanet2019}:

\begin{equation}
    \epsilon(d) \sim \mathcal{N}(0, \sigma_{1}^2 + d \sigma_{2}^2)
\end{equation}

where $\epsilon(d)$ is the error distribution. The approach assumes that the errors follow a Gaussian distribution and the variance is linearly correlated with $d$. The parameters $\sigma_1^2$ and $\sigma_2^2$ are first estimated by minimizing the negative log-likelihood on a calibration set and then used to calculate the predicted standard deviation as an estimate of uncertainty. 

\subsection{Conformal Prediction}
\label{sec:cp}

Conformal prediction is a distribution-free UQ approach with guaranteed finite sample coverage\cite{Angelopoulos2021,Romano2019ConformalizedRegression,Lei2016Distribution-FreeRegression}, ensuring calibration. Many heuristic uncertainty metrics, such as the standard deviation approximated by the ensemble method or the latent distance metric, usually assume the error distribution to be Gaussian. In contrast, CP only assumes that the inputs and outputs are independent and identically distributed (i.i.d.) variables and leverages quantile regression to relax distributional assumptions. Although still generally considered a strong assumption, i.i.d. is ubiquitous in statistical ML schemes and considered a mild assumption compared to the assumption of normally distributed errors. 

The most desirable property of CP in ML regression applications is the guaranteed finite sample coverage, or finite sample validity: 

\begin{equation}
    \mathbb{P}\{Y_{n+1} \in C_{n}(X_{n+1})\} \geq 1 - \alpha,
    \label{eqn:cp_coverage}
\end{equation}

where $(X_{n+1}, Y_{n+1})$ is a new set of explanatory and response variables, respectively, $C_{n}$ is the prediction set based on previous observations $(X_i, Y_i)$ for $i \in 1, ..., n$, and $\alpha$ is a user-defined hyperparameter that sets the desired confidence level. The framework of CP as a generic approach guarantees the probability of the sets containing the ground truth on a new observation, given the prediction sets based on previous observations, is fixed to $1 - \alpha$. See Angelopoulos and Bates\cite{Angelopoulos2021} for the detailed theorem and proof of the coverage property. 

\begin{figure}[htb!]
	\centering
    \includegraphics[width=0.9\linewidth]{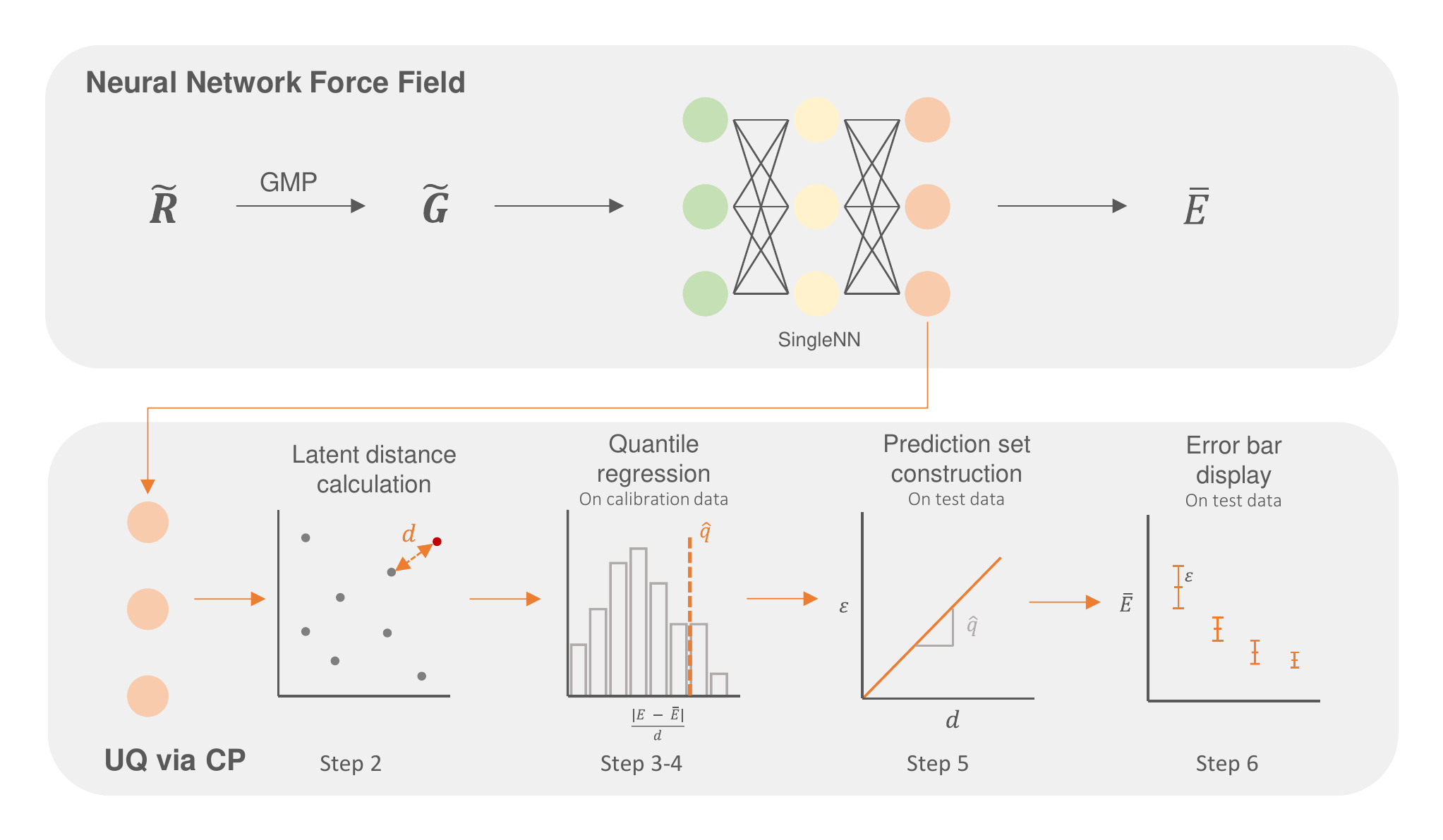}
    \caption{Schematic of CP's workflow on a trained neural network force field. The top block shows the general workflow of an HDNN: generating the local features from Cartesian coordinates, feeding into a neural network model, and obtaining the predicted energy of a given chemical system. In an SNN model, green-colored nodes are the input features, and orange-colored nodes are the last-layer latent features. The latent features are then fed into the bottom block denoting the CP method. The flow includes k-nearest distance computation, quantile regression on calibration data, and prediction on a new data point, following the protocol described in Method \ref{sec:cp} section.}
    \label{fig:workflow_schematic}
\end{figure}

Based on the guaranteed coverage property, CP can convert an arbitrary heuristic notion of uncertainty to a statistically rigorous and calibrated one. As shown in Figure \ref{fig:workflow_schematic}, we applied CP to neural network potentials using the recipe as follows:

\begin{enumerate}
    \item Uniformly sample a fraction (percent of calibration data) of the test data as calibration data.
    \item Calculate the distance between the calibration data and training data in the original feature space or neural network latent space as the heuristic notion of uncertainty. The distance specifically is the Euclidean distance averaged over k-nearest neighbors (num. nearest neighbors). See details on distance definitions below in Section. \ref{sec:distance}. 
    \item Calculate the ratio of the neural network potential residuals over the heuristic distances $\frac{|E - \hat{E}|}{d(X)}$ as the score function. 
    \item Use quantile regression to compute the $\frac{(n+1)(1 - \alpha)}{n}$-th quantile of the score function on calibration data denoted as $\hat{q}$. In this formalism, $\hat{q}$ can be thought of as the scaling factor between the computed distance heuristics and the absolute residuals. 
    \item Apply the $\hat{q}$ to new observations in test data by multiplying with the computed distances to obtain uncertainty, $\epsilon = \hat{q} \times d$. This step is equivalent to calculating $C_{n}(\tilde{G}_{n+1})$ as in Eqn.
    \ref{eqn:cp_coverage}. 
    \item Finally, the error bars are displayed as $\pm \epsilon$. With every prediction, the prediction set is defined as $[\hat{E} - \epsilon, \hat{E} + \epsilon]$. 
\end{enumerate}

The above protocol is implemented in the Python package, \texttt{AmpTorch}. In the SI, we enclose an example script as Listing \ref{lst:code_cp} on how to make uncertainty predictions, and resources for the Python module \texttt{AmpTorch} for a corresponding wrapper and example script. The \texttt{ConformalPrediction} class first takes in the residuals and corresponding heuristic uncertainty metrics on the calibration data, performs quantile regression with user-defined hyperparameter $\alpha$, and finally returns the product of $\hat{q}$ and heuristic uncertainty metrics as the statistically rigorous uncertainty. This formalism of CP does not restrict the heuristic uncertainty metrics to the distance metrics such as feature or latent space distances discussed primarily in this work. CP can also be incorporated with other scalar uncertainty estimates, such as approximated standard deviations from the ensemble method, in place of the heuristic distance metrics effectively as a re-calibration scheme. 

While the guaranteed finite sample coverage is an appealing property, conformal prediction does not provide any guarantees on the sharpness of the uncertainty estimation, especially with an inadequate heuristic notion of uncertainty. Therefore, the goal of CP is to find a heuristic notion of uncertainty that reliably has large values for high uncertainty and low values for low uncertainty. If this requirement cannot be satisfied, the finite sample coverage would still be valid, and thus the uncertainty would be calibrated. However, the error bars on the prediction sets would be unnecessarily large.

\subsection{Distance in the feature or latent space}
\label{sec:distance}

A good choice of heuristic uncertainty notion is essential for conformal prediction to have sharp prediction sets. As pointed out in Section \ref{sec:cp}, any scalar uncertainty heuristic can be a feasible candidate in CP's score function, such as estimated variances from the ensemble approach. However, the estimated variances could become zero for out-of-domain data points with the ensemble approach\cite{Liu2019MolecularCompounds}. Another class of scalar uncertainty heuristics includes distance metrics that measure the dissimilarity between a new data point to the training data, including distances in the feature space. This work focuses on the latter, specifically the Euclidean distance in the neural network's latent space, as first published by Janet et al.\cite{PaulJanet2019} To calculate the distance heuristics, we first compute the image-wise representations in neural network's last-layer latent space or feature space by averaging over all atoms for calibration, test and training data. Then we obtain the Euclidean distances between a calibration or test data point to the k-nearest neighbors from training data in feature or latent representations via the \texttt{pykdtree.kdtree.KDTree} algorithm, and use the average Euclidean distance as the final uncertainty distance heuristic for every data point. We explored two more distance metrics in addition to Euclidean distances on the QM9 dataset in Table \ref{tab:distance_metrics} and did not observe signfinicant differences in calibration or sharpness. Both feature and latent distance metrics are heuristic uncertainty metrics that meet the criterion of conformal prediction: as the heuristic increases, the model uncertainty is expected to increase. 

Distance in the latent space uses the trained neural network model itself as an effective feature engineering tool. It, therefore, introduces no additional costs to model training and evaluation compared to the widely adopted ensemble method and the dropout method. In the case of other types of ML models a similar approach can be used directly with the feature space. Some comparisons to the direct feature space approach are included in this work, revealing that the latent space generally provides sharper UQ estimates, especially at high confidence, and can improve computational efficiency by reducing the dimensionality of the feature vector.

\section{Results}

\subsection{Defining the confidence level in CP}

\begin{figure}[htb!]
	\centering
    \includegraphics[width=0.9\linewidth]{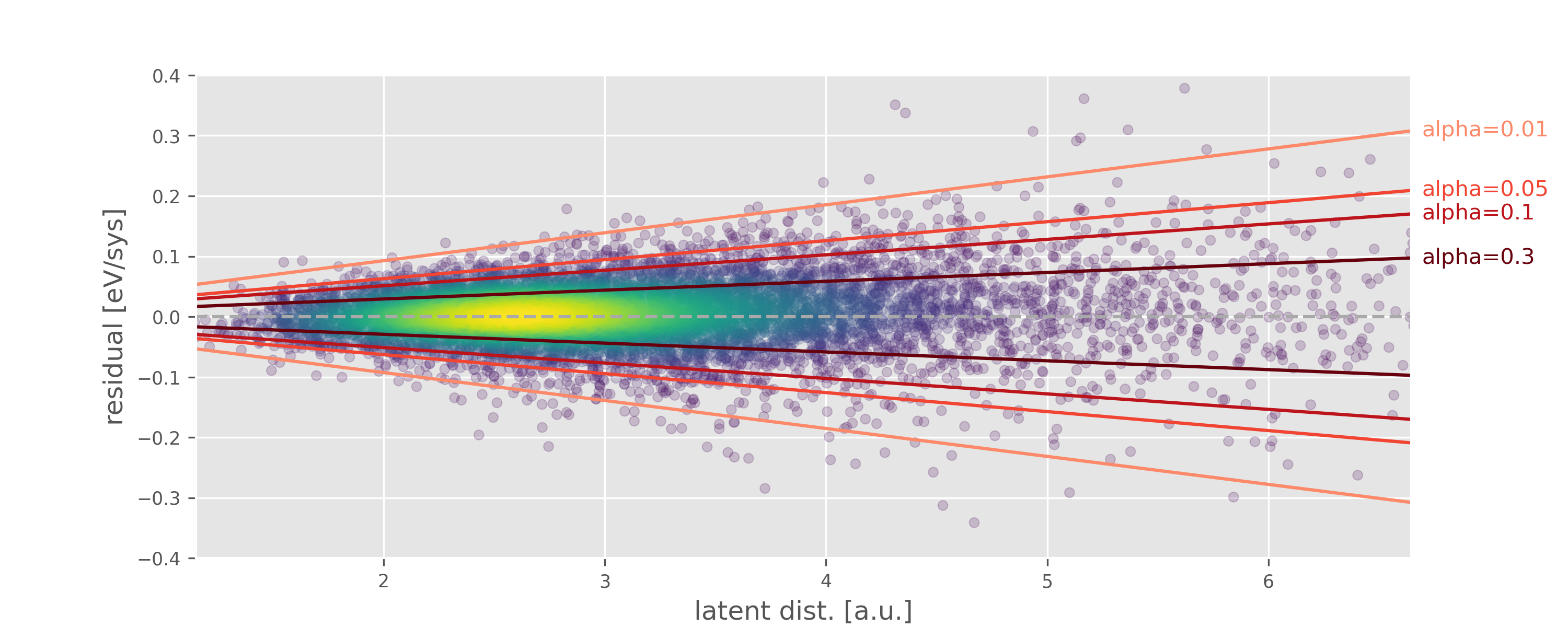}
    \caption{Prediction sets with confidence level $(1-\alpha)100\%$ as a function of distances based on $\hat{q}$ from calibration data shown as pairs of upper and lower lines for different $\alpha$'s. Hyperparameter $\alpha$ in CP defines the confidence level and changes the width of the prediction sets. In the y-axis, residuals are the difference between ground truths and predictions. The x-axis is the Euclidean distances between a test data point to training data in the last-layer of the latent space. Between the two lines, there is $\sim (1-\alpha)100\%$ of total data given i.i.d. assumption. Data points are from QM9 dataset. The color of the dots (from purple to yellow) denotes density in close proximity by KDE analysis. The brighter the color is, the more densely populated points are. }
    \label{fig:cp_alphas}
\end{figure}

The valid confidence level for the CP model as a result of the finite sample coverage property is defined by $(1-\alpha)100\%$.  Figure \ref{fig:cp_alphas} provides a graphical illustration of the result of CP applied to the QM9 dataset, where $\alpha$ changes from 0.01, 0.05, 0.1 to 0.3 (color-coded) and associated expected confidence levels are 99\%, 95\%, 90\% and 70\% respectively. Prediction sets (i.e. error bars) for the calculated distance in latent space are defined by the two lines of the respective color for a given $\alpha$. The prediction sets get sharper as $\alpha$ increases and the confidence level decreases, and the percentage of residuals outside the prediction set bound is approximately equal to $100\alpha \%$. The visualization in Fig. \ref{fig:cp_alphas} provides insight into the dependence of the residuals on the distance metric, the size of error bar as a function of $\alpha$ and the distance metric, and the frequency and magnitude of residuals outside the predicted error bars as a function of the distance metric.

\subsection{Comparison of UQ methods}

\subsubsection{Calibration}

\begin{figure}[htb!]
	\centering
	\includegraphics[width=\textwidth]{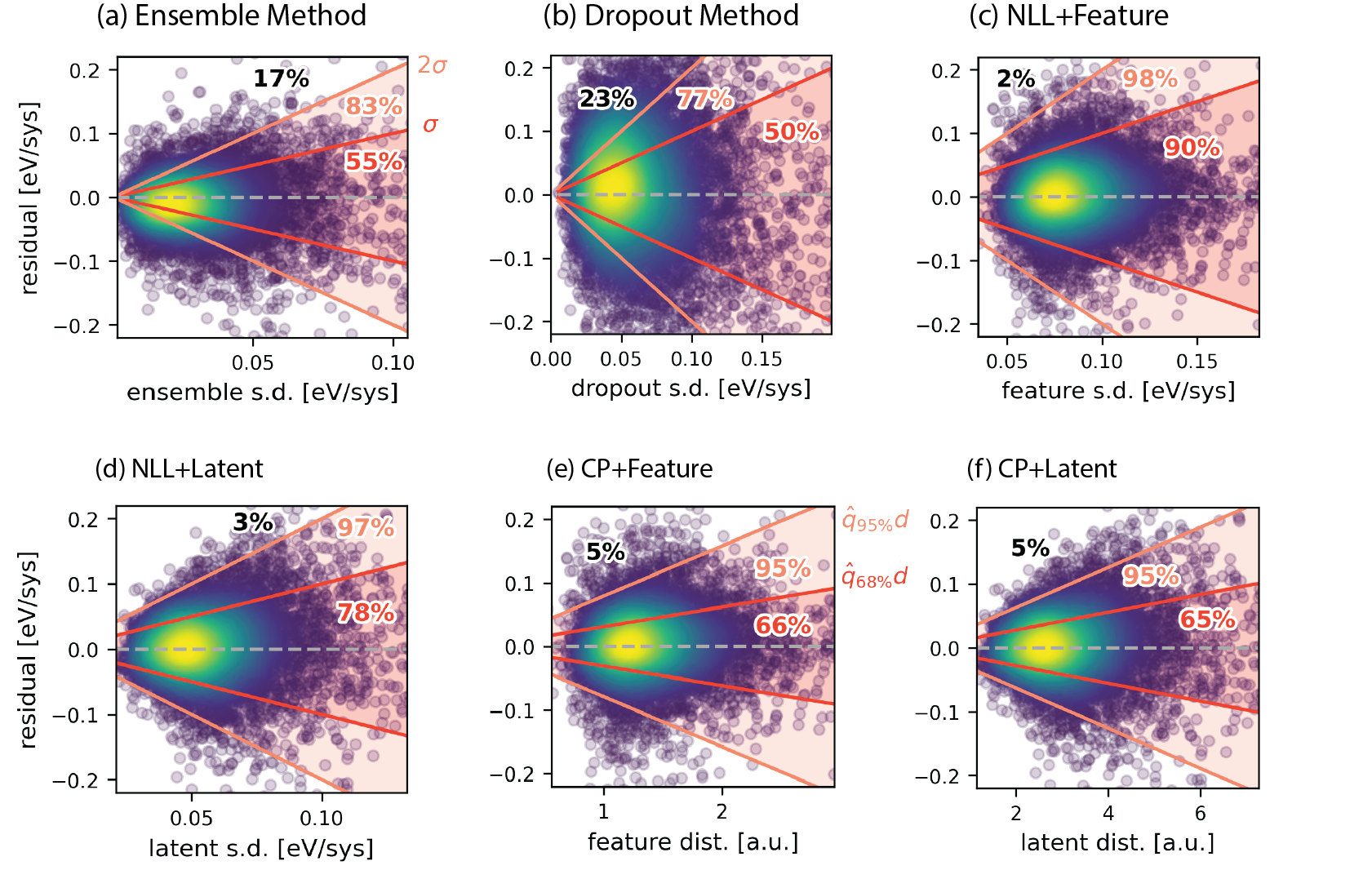}
    \caption{Scattering of residuals vs. uncertainty heuristics for different UQ methods on QM9 dataset. (a,b,c,d) The region bounded by the upper and lower red/orange lines is where test data are covered within one/two standard deviation(s). The observed percents of test data covered within one, two, and outside two standard deviation(s) are annotated in bold red, orange, and black fonts respectively. Assuming normally distributed errors, the expected values for percents are 68\%, 95\% and 5\%. (e,f) The region bounded by upper and lower red/orange lines is the prediction set of expected $(1-\alpha)100\%=68\%$/$95\%$ confidence level. These two specific numbers are chosen to draw resemblance to the above approaches that assume normally distributed errors. The observed percents of test data covered within the 68\% and 95\% and outside 95\% prediction sets are annotated in bold, red, orange, and black fonts respectively. }
    \label{fig:method_comparison_qm9}
\end{figure}

In this section, we compare 6 UQ methods with the residuals plotted against the respective uncertainty heuristics and their expected confidence levels to visually assess model performance and calibration. The UQ methods include:

\begin{enumerate}
    \item the ensemble method (Figure \ref{fig:method_comparison_qm9}a) constructed with 4 different atomistic neural network structures, all trained on the same data set
    \item the dropout method (Figure \ref{fig:method_comparison_qm9}b) constructed with 4 different atomistic neural networks at \texttt{dropout\_rate = 0.2} trained on the same data set\cite{Tran2020MethodsPredictions}
    \item distances in the feature space with negative log-likelihood (NLL) estimation\cite{PaulJanet2019} (Figure \ref{fig:method_comparison_qm9}c)
    \item distances in the latent space with NLL estimation\cite{PaulJanet2019} (Figure \ref{fig:method_comparison_qm9}d)
    \item distances in the original GMP feature space with CP (Figure \ref{fig:method_comparison_qm9}e)
    \item distances in the last-layer latent space with CP (Figure \ref{fig:method_comparison_qm9}f)
\end{enumerate}

The ensemble, dropout, and NLL methods assume that the errors follow Gaussian distributions. Therefore the presumed confidence levels for the coverage within one and two standard deviations are 68\% and 95\%. We plot the scattering of model residuals over different uncertainty heuristics on test data of QM9 dataset in Figure \ref{fig:method_comparison_qm9}. The numbers of expected and observed confidence levels and the difference are tabulated in Table \ref{tab:confidence_levels_qm9}. The observed confidence levels are off for the ensemble method by +13\%/+12\% and for the dropout method by +18\%/+18\% referenced to the expected 68\% and 95\% confidence levels. 
For NLL methods, NLL+latent method conforms to the expected confidence levels better (+10\%/-2\%) than NLL+feature method (+22\%/-3\%), consistent with Janet et al.\cite{PaulJanet2019} On the contrary, both CP methods (CP+feature and CP+latent) produce the observed confidence levels within 3\% to the presumed values, regardless of the uncertainty heuristics chosen. We performed the same analysis on dataset MD17-Aspirin, and confirmed the findings are qualitatively equivalent, with the CP method being the best calibrated (Figure \ref{fig:method_comparison_md17} \& Figure \ref{fig:calibration_curve_md17}).

\begin{figure}[htb!]
	\centering
    \begin{subfigure}[t]{0.3\textwidth}
		\centering
        \subcaption[short for lof]{Ensemble method}
        \includegraphics[width=\linewidth]{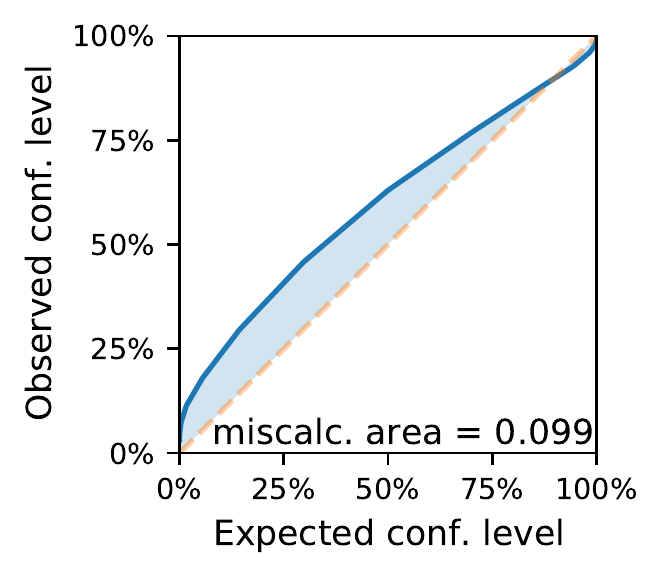}
        \label{fig:ensemble_calibration_curve}
	\end{subfigure}\hspace{0.005\textwidth}
    \begin{subfigure}[t]{0.3\textwidth}
		\centering
        \subcaption[short for lof]{Dropout method}
        \includegraphics[width=\linewidth]{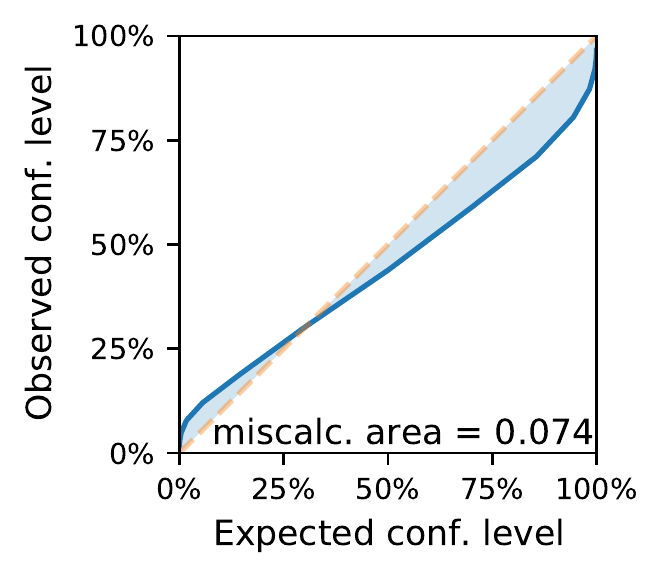}
        \label{fig:dropout_calibration_curve}
	\end{subfigure}\hspace{0.005\textwidth}
    \begin{subfigure}[t]{0.3\textwidth}
		\centering
		\subcaption[short for lof]{NLL+feature}
        \includegraphics[width=\linewidth]{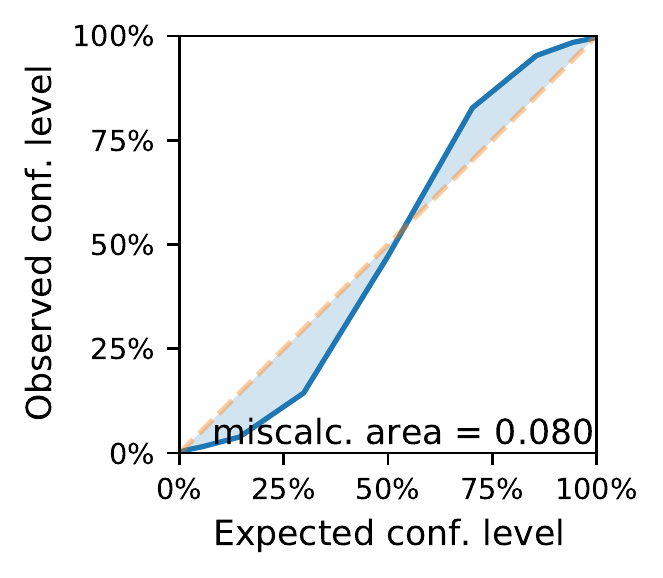}
        \label{fig:nll_feature_calibration_curve}
	\end{subfigure}\hspace{0.005\textwidth}%
    \begin{subfigure}[t]{0.3\textwidth}
		\centering
		\subcaption[short for lof]{NLL+latent}
        \includegraphics[width=\linewidth]{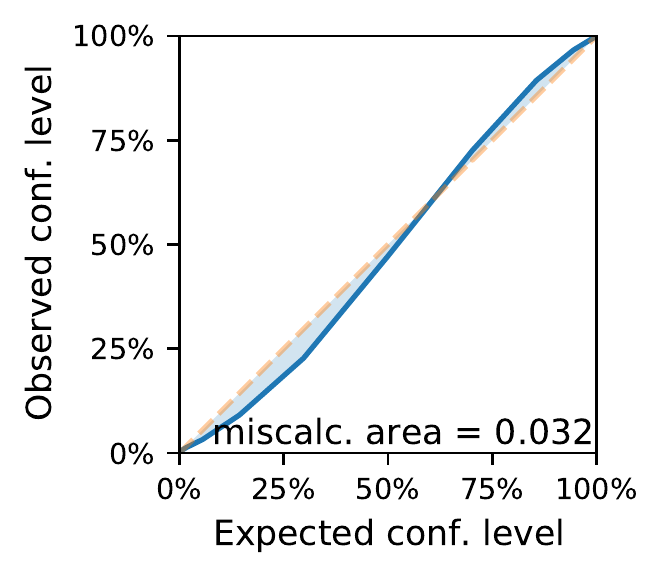}
        \label{fig:nll_latent_calibration_curve}
	\end{subfigure}\hspace{0.005\textwidth}%
    \begin{subfigure}[t]{0.3\textwidth}
		\centering
		\subcaption[short for lof]{CP with feature distances}
        \includegraphics[width=\linewidth]{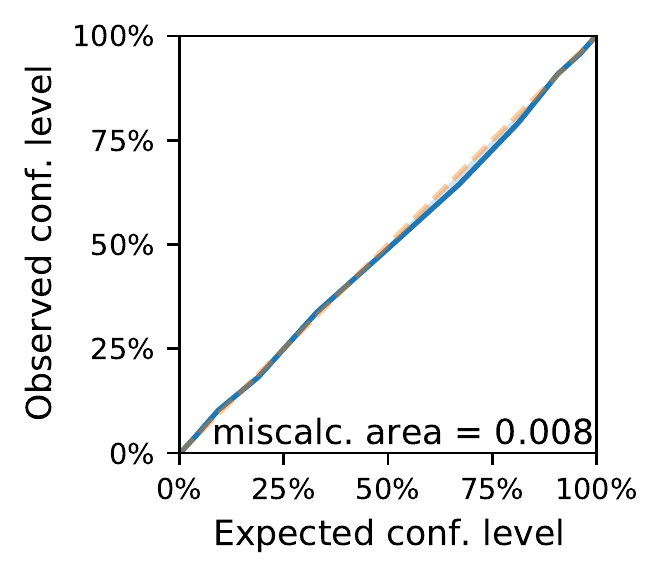}
        \label{fig:cp_feature_calibration_curve}
	\end{subfigure}\hspace{0.005\textwidth}%
    \begin{subfigure}[t]{0.3\textwidth}
		\centering
		\subcaption[short for lof]{CP with latent distances}
        \includegraphics[width=\linewidth]{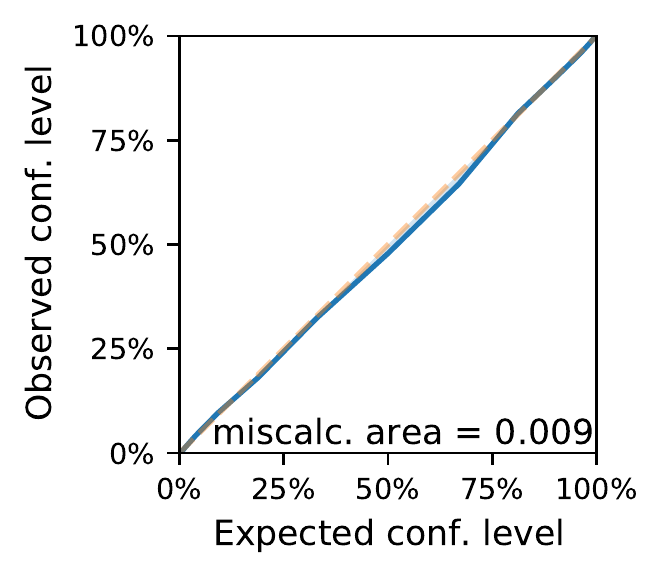}
        \label{fig:cp_latent_calibration_curve}
	\end{subfigure}\hspace{0.005\textwidth}%
    \caption{Calibration curves of different UQ methods on QM9 dataset. The orange dashed line ($y=x$) is the ideal case where the observed confidence level always matches the expected confidence level. For ensemble, dropout, and NLL methods, the distribution is Gaussian. For CP methods, the expected confidence level is $(1-\alpha)100\%$. The solid blue line is the actual observed confidence level on test data. The area between the solid blue line and the orange dashed line indicates the frequency the method is miscalibrated. At the bottom of every plot prints the values for miscalibrated areas. }
    \label{fig:calibration_curve_qm9}
\end{figure}

To better compare the calibration, the calibration curves of different UQ methods are plotted in Figure \ref{fig:calibration_curve_qm9} with various observed and expected confidence levels following the approach in prior publications\cite{Tran2020MethodsPredictions, Pernot2022TheChemistry}. The ensemble and the dropout methods have relatively high miscalculated areas, suggesting that they are not well calibrated. When used with the latent space distances, the NLL method is more calibrated compared to feature space distances\cite{PaulJanet2019}. CP methods display the best calibration of all the methods by having the miscalculated area close to 0.01. The fact that all methods relying on the assumption of normally distributed errors are not as well-calibrated as distributional-free CP methods indicates the invalidation of underlying assumption on error distribution, as noted by Pernot\cite{Pernot2022TheChemistry}. We have performed the same analysis on MD17-Aspirin dataset (Figure \ref{fig:method_comparison_md17} \& Figure \ref{fig:calibration_curve_md17}). We observed that NLL methods have similar performance as the CP methods on the MD17-Aspirin dataset, mainly because this dataset contains relatively simple chemical interactions, and the errors mostly follow Gaussian distributions. The calibration analysis further supports the point made by Pernot\cite{Pernot2022TheChemistry} that the parametric errors should not be assumed to be normal without testing, and showcases the advantage and the general applicability of distribution-free CP methods. 

\subsubsection{Sharpness}

In addition to calibration, which ensures the reliability of a UQ method, sharpness is an indicator of the usefulness of a UQ method. Here we compare the calculated sharpness on the QM9 test data for all six methods. The ensemble, dropout, and NLL methods use $sha_{\sigma}$. We use $sha_{68\%}$ for CP methods to draw a close prediction since distribution-free CP methods produce prediction bandwidths given a confidence level instead of the standard deviation. Although it is only meaningful to compare sharpness for calibrated UQ methods\cite{Pernot2022TheChemistry}, we report the calculated sharpness for all methods (Table. \ref{tab:sharpness_qm9}) but focus the discussion only on relatively calibrated ones: NLL+latent, CP+feature, and CP+latent. Both CP methods have smaller $sha$ values, indicating that they are sharper, than the NLL method with latent space distances among the three methods. This is not surprising because NLL+latent is less calibrated than CP methods and has an observed confidence level at 78\%, which is higher than the expected 68\% for one standard deviation coverage assuming the errors follow a Gaussian distribution. The fact that the error distributions are not normal affects not only the level of calibration but also the calculated sharpness. Furthermore, the sharpness of the CP method is better than all methods except the ensemble approach, which is also the least calibrated. The low sharpness of the ensemble approach is related to the fact that it is poorly calibrated and systematically under-estimates the actual error.

\begin{table}[htb!]
    \centering
    \caption{Sharpness comparison. All units are in meV/sys. The sharpness for the ensemble, dropout and NLL methods is reported as the mean of standard deviations ($sha_{\sigma}$) on QM9 test data. The sharpness of CP methods is reported as the mean of the prediction bandwidths of expected confidence level 68\% and 99\% on test data ($sha_{68\%}$, $sha_{99\%}$). The $\downarrow$ symbol indicates that methods with smaller $sha$ values are considered better after they are proven calibrated. The less/least value among calibrated methods is underscored.}
    \vspace{0.5em}
    \begin{tabular}{c c | c c | c c c}
        \hline
        \thead{Method} & $sha_{\sigma} (\downarrow)$  & \thead{Method} & $sha_{\sigma} (\downarrow)$ &  \thead{Method} & $sha_{68\%} (\downarrow) $ & $sha_{99\%} (\downarrow)$ \\
        \hline 
        Ensemble & 30.66 & NLL + feature & 87.25 & CP + feature &  \underline{43.54} & 201.59 \\
        Dropout & 63.54 & NLL + latent & 57.90 & CP + latent & 44.12 & \underline{146.99} \\
         \hline
    \end{tabular}
    \label{tab:sharpness_qm9}
\end{table}

To compare CP+feature and CP+latent, we concur with Janet et al.\cite{PaulJanet2019} that the distances in the latent space are a better heuristic than distances in the original feature space. The CP+feature and CP+latent have a similar miscalibrated area in calibration analysis and similar performance in $sha_{68\%}$ as shown in Table \ref{tab:sharpness_qm9}. However, we further analyzed $sha_{95\%}$ of the two methods, where both methods show a 95\% observed confidence level. In this case, the CP+feature has $sha_{95\%} = 111.30$ meV/sys, while CP+latent has $sha_{95\%} = 99.85$ meV/sys. The difference in sharpness is due to the important observation that high errors happen less frequently at low latent distances (Figure \ref{fig:method_comparison_qm9}f), as opposed to the original feature distances (Figure \ref{fig:method_comparison_qm9}e). Janet et al. reported the same observation\cite{PaulJanet2019}. Using latent space distances ensures the sharpness of the model at high confidence levels due to a better correlation with uncertainty at low latent distances, and the discrepancy in sharpness increases with the confidence level ($sha_{99\%} = 201.59$ meV/sys for CP+feature, $sha_{99\%} = 146.99$ meV/sys for CP+latent). This is an important distinction when trying to obtain sharp prediction sets in physical simulations and active learning schemes where the threshold for the confidence level is typically high. 
Another important reason to favor latent space distances over feature space distances is the better scalability in the latent space, as discussed in Section. \ref{sec:scalability_and_portability}.

\subsection{Sensitivity to model hyperparameters}

In this subsection, we seek to investigate how the model hyperparameters affect the level of calibration, sharpness, and scalability of CP methods. The hyperparameters of the CP model include $\alpha$, whose values determine the confidence of the UQ model to be $(1-\alpha)100\%$, the number of k-nearest neighbors (num. nearest neighbors) in distance calculation, and the percent of test data to be taken as calibration data (per. calib.). 

\begin{figure}[htb!]
	\centering
    \includegraphics[width=\linewidth]{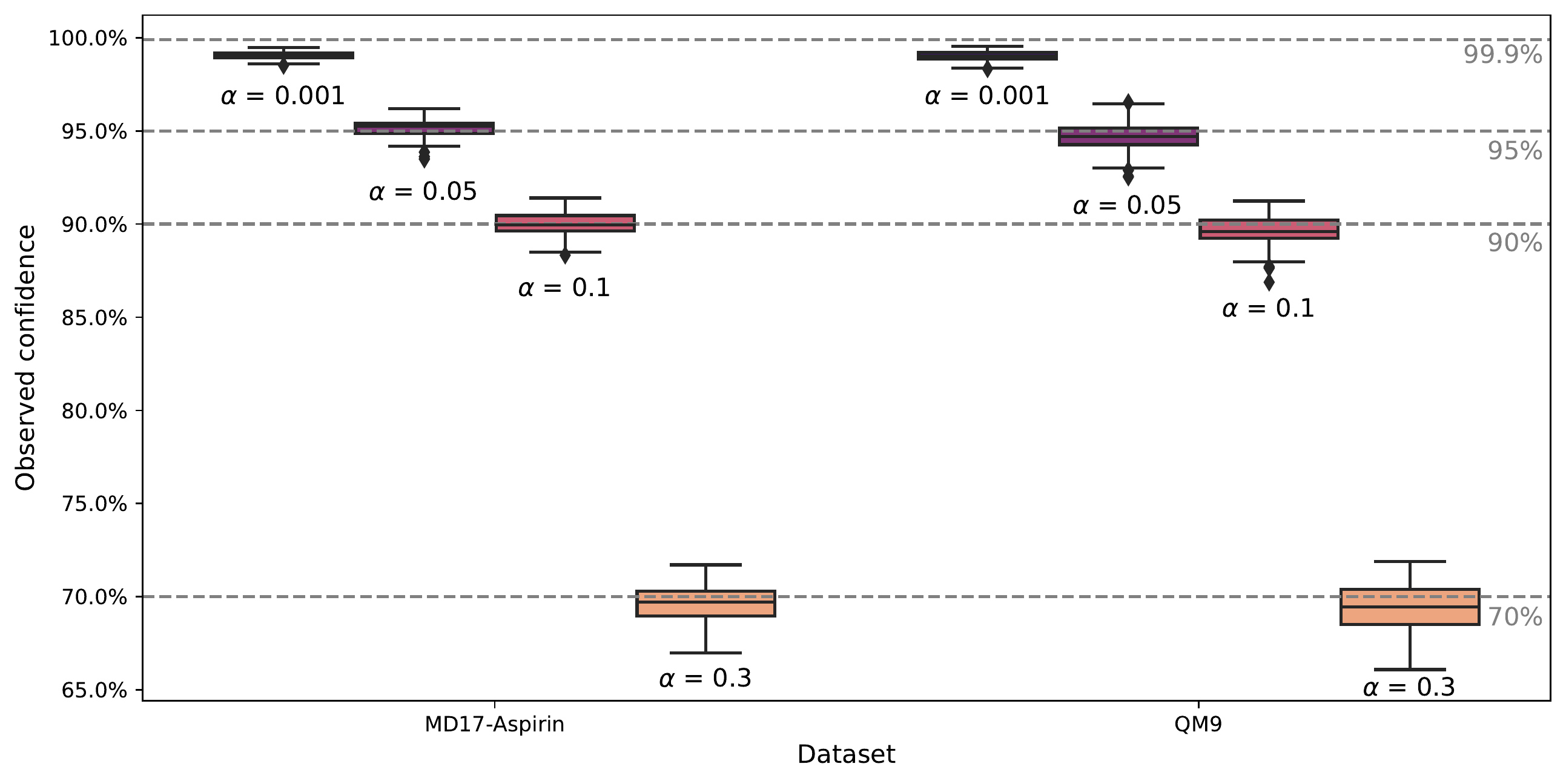}
    \caption{Boxplot of observed confidence levels for datasets QM9 and MD17-Aspirin under various model hyperparameters (models trained with different number of training data, $\alpha$, number of nearest neighbors in distance calculation, random seed for calibration/test split, and percent of allocated calibration data). The observed confidence levels on test data are close to $(1-\alpha)100\%$ (gray dotted lines with labels underneath to the right) irrespective of model hyperparameters as long as the i.i.d. assumption for calibration and test data is valid. }
    \label{fig:box_calibration}
\end{figure}

To study whether the models are calibrated, we plot the observed confidence levels with various choices of hyperparameters for two datasets, QM9 and MD17-Aspirin, in Figure \ref{fig:box_calibration} with expected confidence levels plotted as gray lines for different $\alpha$'s. The boxplot contains all the CP models with different $\alpha$'s, number of nearest neighbors ranging from 2 to 300, percent of calibration data ranging from 5\% to 30\%, random seeds for calibration/test split, and three different neural network force fields, each trained with small (20,000), medium (50,000) or large (120,000) training datasets. Despite the significant variations in datasets, model hyperparameters and even neural network model accuracy, the observed confidence levels are close to $(1-\alpha)100\%$ (maximum deviation of $\sim$3\%), indicating that the calibration property of CP models is insensitive to model these factors. 


We further evaluated the calibration, sharpness, and efficiency as a function of alpha, number of nearest neighbors, and percent of calibration data (QM9: Figs. \ref{fig:calibration_parameter_qm9} \& \ref{fig:sharpness_parameter_qm9}, MD17-Aspirin: Figs. \ref{fig:calibration_parameter_md17} \& \ref{fig:sharpness_parameter_md17}). The results show that calibration and sharpness are relatively insensitive to all parameters, particularly if percent calibration is above 10\% and number of nearest neighbors is greater than 10. The model efficiency is most sensitive to number of nearest neighbors, particularly for large training sets, with evaluation time rising logarithmically with the number of nearest neighbors, with 10 nearest neighbors being recommended for the optimal tradeoff between sharpness and efficiency. The findings are consistent for both QM9 and MD17-Aspirin datasets, so we expect that 10 nearest neighbors and 10\% calibration data are reasonable defaults in most scenarios. We note that these recommendations may change in the limit of very small datasets ($<$ 100-1000 data points), where the small number of data points in the calibration set may cause the results to be more sensitive to exactly which points are selected, and where the number of nearest neighbors becomes a significant fraction of the entire dataset.

\subsection{Analysis on out-of-distribution set}
\label{sec:ood}

\begin{figure}[hbt!]
	\centering
    \includegraphics[width=0.7\linewidth]{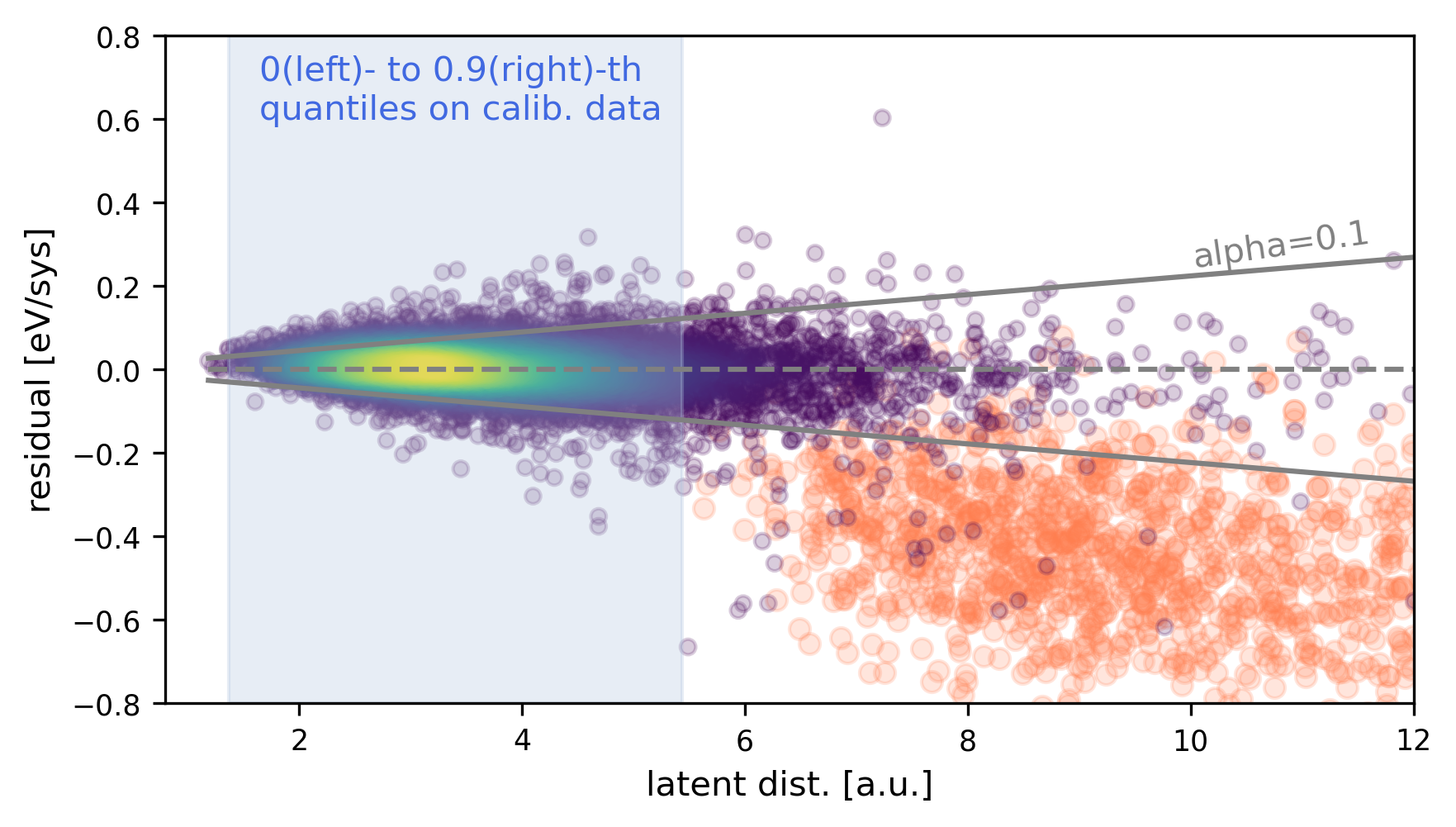}
    \caption{Out-of-distribution analysis on QM9. Yellow-to-purple dots are test data points from the same distribution as calibration data. Orange dots are out-of-distribution data points. Empirical cutoffs from 0th- to 0.9th- quantiles computed from latent distances of the calibration data are plotted as the left and right bounds of the blue box, and the cutoffs separate out the out-of-distribution data points. }
    \label{fig:out_of_distribution}
\end{figure}

In applications such as materials discovery and active learning, it is expected (and even desired) that models should move beyond the distribution of the initial training set to identify novel regions of the potential energy surface or the material search space\cite{Shaidu2021ACarbon,Meredig2018CanDiscovery}. However, this presents a challenge since nearly all UQ techniques, including CP, rely on the i.i.d. assumption. To establish the limitations of the CP+latent technique, this section explores how the CP model performs on out-of-distribution test data points. Ideally, we want the UQ method to predict higher uncertainty for out-of-distribution data than for in-distribution data. Here, we test an extreme case where the model is asked to extrapolate to an unseen element, and evaluate the resulting uncertainty predictions. For this analysis the QM9 dataset is divided into two separate datasets, one dataset named \texttt{F-} consisting of molecules made up only of C, H, O, N without fluorine (F) atoms, and another \texttt{F+}, with all molecules that contain F atoms. We trained a neural network model on dataset \texttt{F-} and constructed the CP model with a calibration set homogeneous to \texttt{F-}. We then estimated the uncertainty using the CP+latent method for two test sets in Figure \ref{fig:out_of_distribution}. Yellow to purple dots indicate the test set drawn from the same distribution as the training and calibration data, \texttt{F-}. Orange dots indicate the out-of-distribution test set, \texttt{F+}, with the observed confidence level at 26\% (off by 64\%). It is clear that CP can no longer guarantee the sample coverage for out-of-distribution data with respect to the expected 90\% confidence level due to a clear violation of the i.i.d. assumption. However, the distributions for \texttt{F-} and \texttt{F+} sets also have a very different distribution of latent distances, with the average and standard deviation of latent distances being 3.78 (with a standard deviation of 1.38) for \texttt{F-} and 13.85 (with a standard deviation of 7.88) for \texttt{F+}. The poor calibration on out-of-distribution data is consistent with the prior observation that higher errors happen less frequently at lower latent distances. As a result, we propose using a quantile cutoff on the latent distances of the calibration data as an \textit{ad hoc} remedy to identify the out-of-distribution data points. In Figure \ref{fig:out_of_distribution}, we plot the range of latent distances of calibration data up to its 90\%-quantile in the blue block. A clear separation between the in-distribution and out-of-distribution points shows that empirical cutoffs can effectively distinguish whether the CP method is confident about its estimation of uncertainty based on the known latent distance distribution. The generality of this approach is not guaranteed, and it may prove challenging to define the cutoff \textit{a priori}, but this example provides evidence that a latent distance cutoff is an effective strategy to identify out-of-distribution data in at least some cases. To further test the applicability of the CP+latent method in an active learning scheme, we iteratively trained models with the most uncertain OOD F+ molecules and tested the model accuracy in MAE, calibration in observed confidence, and average uncertainty in sharpness in Figure \ref{fig:ood_active_learning} (and corresponding parity plots in Figure \ref{fig:active_learning_parity}). Both MAE and sharpness improve, suggesting the model gets more accurate and confident about predictions on F+ molecules with more supplemented data. The observed confidence level also gets closer to the expected value, but rigorous calibration is not expected in general due to a violation of the i.i.d. assumption. This challenge will exist for any UQ scheme that relies on the i.i.d. assumption, and the fact that the CP+latent method improves as more data is added suggests that it has practical utility in active learning schemes, although the details of the performance will likely vary depending on the specifics of the data set and the sampling scheme utilized.

\subsection{Scalability and portability}
\label{sec:scalability_and_portability}

Besides quality metrics such as calibration and sharpness, scalablity determines if a UQ method can be applied to complex models and large datasets, while portability refers to the ability to apply the approach to pre-trained models of various architectures. Large datasets with complex neural network models are one of the most promising use cases for the CP+latent approach, since no re-training is needed and the neural network model can function as a feature engineering and dimension reduction tool to make the overall uncertainty estimation more cost-effective. The computational bottleneck of the CP method with distance metrics is the KDTree algorithm used to compute the distances to nearest neighbors in the neural network latent space, whose time complexity is approximately $O(D N \log N)$ where $D$ is the number of dimensions and $N$ is the number of points\cite{Camerini1980BuildingTime}. In Figure \ref{fig:scalability_latent_dimension}a), the computational costs for CP method with latent distances can be reduced by as much as 2 orders of magnitude the latent dimension is reduced from 64 to 8 using a funnel neural network architecture. To benchmark the timing of the CP method, we calculate the ratio between the time for distance calculations plus quantile regression (CP+latent) with varying number of latent dimensions and the time required for the forward pass of the neural network (not including time required for fingerprint calculation). This ratio is relevant to compare across different UQ techniques, such as ensembles and dropout networks, since both types require multiple forward passes ($\sim$10-100) to quantify the uncertainty \cite{Wen2020UncertaintyPotentials,Musielewicz2022FINETUNA:Simulations}. In Table \ref{tab:scalability_latent_dimension}, the CP+latent method requires less time than two forward-passes even with the highest number of latent dimensions tested (64). Figure \ref{fig:scalability_latent_dimension}b) and c) shows that the model accuracy and sharpness also tend to improve slightly with a lower latent dimensions (16 and 32), likely benefiting from the funnel-shaped architecture of the corresponding networks. However, as the latent dimension is decreased further the sharpness decreases slightly (at latent dimensions of 8 and 16) and the accuracy also decreases somewhat (at latent dimension of 8). Nonetheless, these decreases are relatively minor ($\sim$0.002 eV), and suggest that funnel-shaped networks are promising for both improving model accuracy and efficiency of UQ using CP.


\begin{table}[htb!]
    \centering
    \caption{Time costs of CP method with latent distances with varying numbers of dimensions in the latent space on QM9 dataset. Reference forward-passing time is the amount of time to pass the loaded representation of a new data point through the neural network model, not including time for generating fingerprints. The ratio to forward-passing (passing the calculated representation through the transformation defined by neural network model) represents the additional cost for UQ estimation and enables comparison to other techniques.}
    \vspace{0.5em}
    \begin{tabular}{c c c c }
    \hline
        \thead{Neural network \\ architecture} & \thead{Latent \\ dimension} & \thead{Time for CP \\ (ms/image)} & \thead{Ratio to \\ Forward-passing} \\ \hline \hline
        [128,64,32,16,8] & 8 & 0.0338 & 1.14\% \\ \hline
        [128,64,32,16,16] & 16 & 0.3057 & 9.94\%  \\ \hline
        [128,64,32,32,32] & 32 & 0.8536 & 27.00\% \\ \hline
        [64,64,64] & 64 & 4.4968 & 185.0\%  \\ \hline
    \end{tabular}
    \label{tab:scalability_latent_dimension}
\end{table}

\begin{figure}[htb!]
	\centering
    \begin{subfigure}[t]{0.33\textwidth}
		\centering
		\subcaption[short for lof]{SingleNN}
        \includegraphics[width=\linewidth]{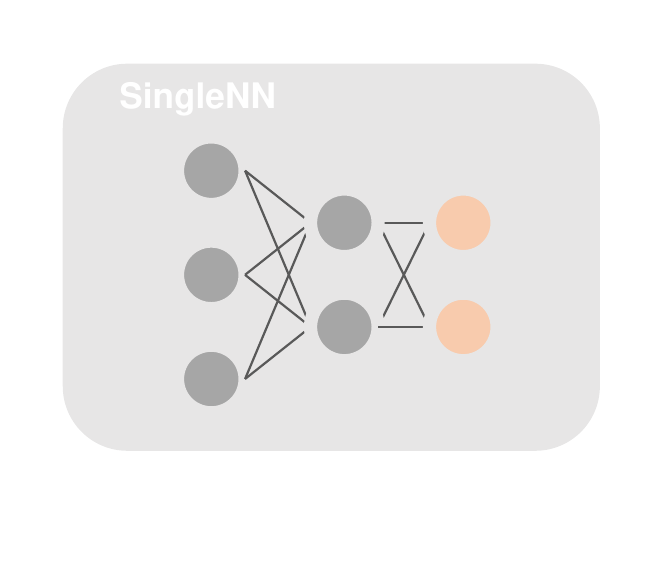}
        \label{fig:singlenn_funnel_schematic}
	\end{subfigure}
    \begin{subfigure}[t]{0.3\textwidth}
		\centering
		\subcaption[short for lof]{NLL+latent in SingleNN}
        \includegraphics[width=\linewidth]{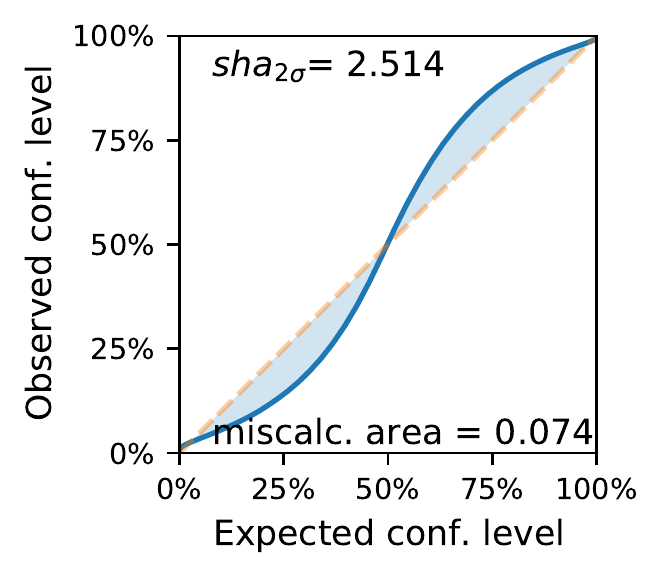}
        \label{fig:singlenn_nll_latent_cc_oc20}
	\end{subfigure}
    \begin{subfigure}[t]{0.3\textwidth}
		\centering
		\subcaption[short for lof]{CP+latent in SingleNN}
        \includegraphics[width=\linewidth]{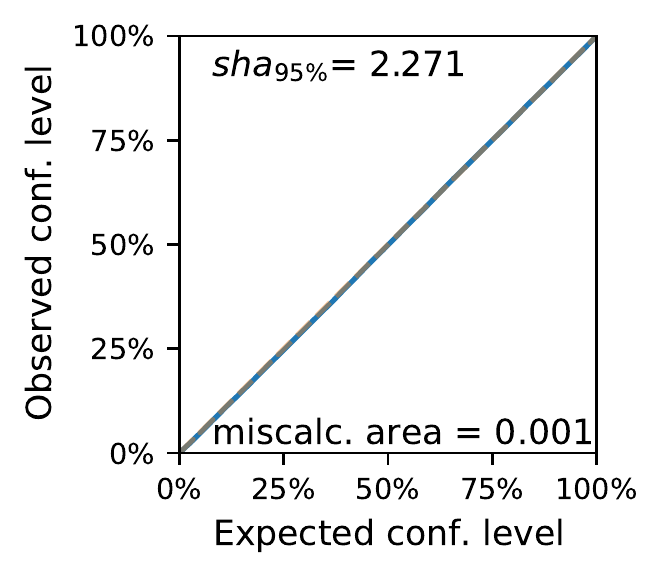}
        \label{fig:singlenn_cp_latent_cc_oc20}
	\end{subfigure}
    \begin{subfigure}[t]{0.33\textwidth}
		\centering
		\subcaption[short for lof]{GemNet-OC}
        \includegraphics[width=\linewidth]{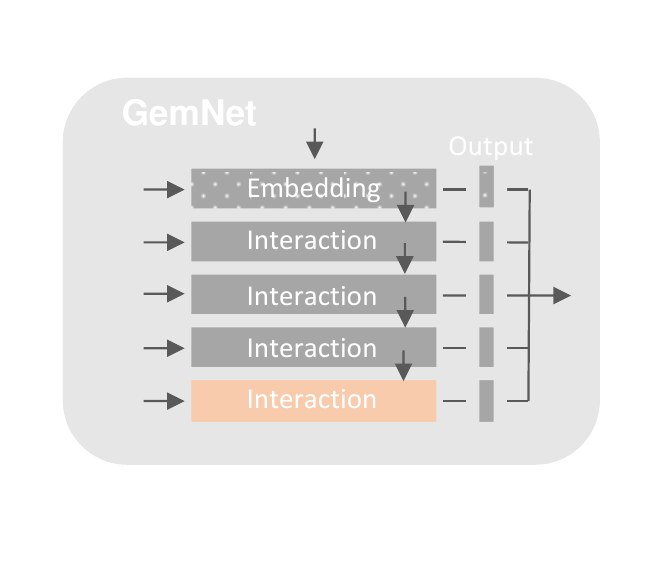}
        \label{fig:gemnet_schematic}
	\end{subfigure}
    \begin{subfigure}[t]{0.3\textwidth}
		\centering
		\subcaption[short for lof]{NLL+latent in GemNet-OC}
        \includegraphics[width=\linewidth]{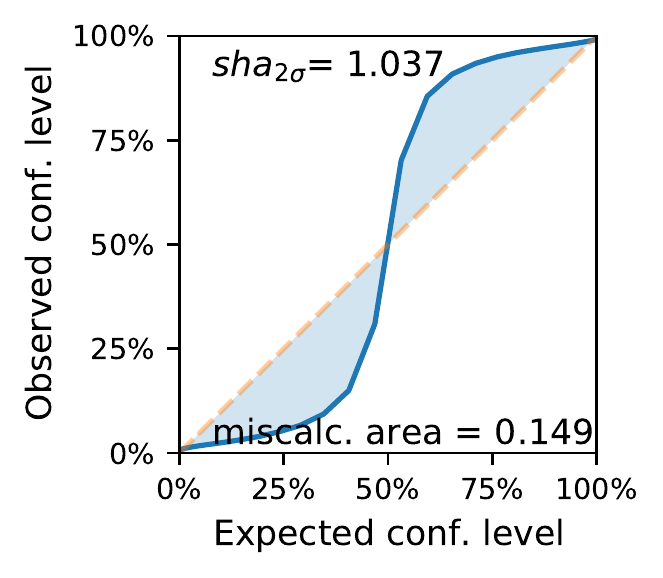}
        \label{fig:gemnet_nll_latent_cc_oc20}
	\end{subfigure}
    \begin{subfigure}[t]{0.3\textwidth}
		\centering
		\subcaption[short for lof]{CP+latent in GemNet-OC}
        \includegraphics[width=\linewidth]{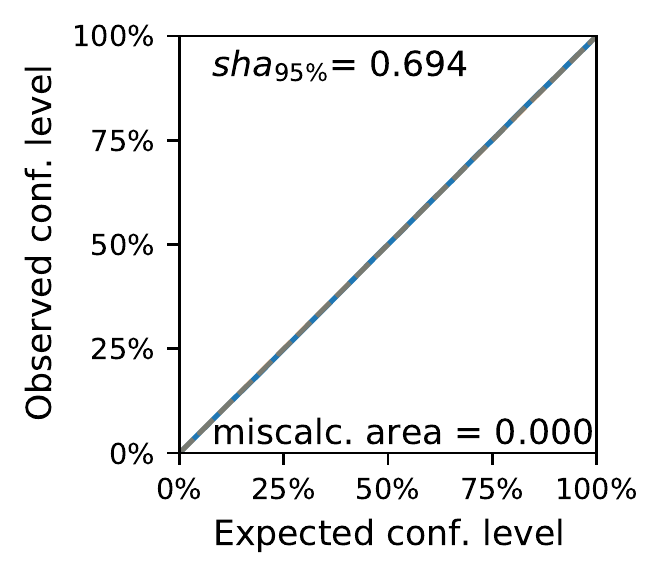}
        \label{fig:gemnet_cp_latent_cc_oc20}
	\end{subfigure}
    \caption{Calibration curves of NLL+latent and CP+latent methods applied to SNN and GemNet-OC models. (a) Schematic of a funnel-shaped SNN. Nodes in the neural network's last latent layer are colored orange and used as the latent space to compute distances in CP between a new data point and test data. (d) Schematic of GemNet-OC, an example of a GNN. The last interaction block is colored orange and used as the latent space for distance computation in CP. Orange nodes/block correspond(s) to the input of the bottom block in Figure \ref{fig:workflow_schematic}. (b,c,e,f) Calibration curves of two distance heuristic methods on two different neural network models trained with OC20 dataset. The orange dashed line ($y=x$) is the ideal case where the observed confidence level always matches the expected confidence level from a Gaussian distribution in the case of NLL or as $(1-\alpha)100\%$ in the case of CP. The solid blue line is the actual observed confidence level on test data. The area between the solid blue line and the orange dashed line indicates the frequency the method is miscalibrated. The bottom of every plot shows the values for miscalibrated areas. Sharpness in unit of eV/sys at 2 $\sigma$'s/ 95\% is printed for NLL methods/CP methods. During calibration step, training data for the UQ methods are 1M data points uniformly selected at random from OC20 S2EF dataset. }
    \label{fig:calibration_curves_oc20}
\end{figure}

Finally, we demonstrate the portability of the CP+latent UQ method by applying it to two pre-trained neural network models, including a modified 2nd generation neural network model, SNN (Figure \ref{fig:singlenn_funnel_schematic}), and a state-of-the-art graph convolution neural network model, GemNet-OC (Figure \ref{fig:gemnet_schematic}). We aim to demonstrate the portability of the methods to complex neural network models trained on large ($>$1M) datasets such as OC20. In the case of SNN, we utilize a pre-trained model based on 91 GMP features with a [128,64,64] neural network architecture that was trained on 1M data points randomly sampled from the OC20 S2EF dataset. The pre-trained GemNet-OC model has 264 dimensions in the interaction blocks and is trained on the entire 134M OC20 dataset. Ensemble and dropout methods require re-training multiple models, resulting in significant computational costs, so no direct comparison is performed. In addition, the number of input features (91) for the SNN model is higher than the the dimension of last-layer latent space, and there is no direct way to sample the feature space of the GemNet-OC model since it uses molecular graphs as inputs. Hence, we focus only on the comparison between the NLL+latent method and CP+latent method for SNN and GemNet-OC models. 

The results of the calibration for both model types are shown in Figures \ref{fig:calibration_curves_oc20} and \ref{fig:method_comparison_oc20}, revealing that the CP+latent method is well-calibrated while the NLL+latent method has larger mis-calibration areas compared to CP+latent method for both neural network models (0.07 for SNN model and 0.15 for GemNet-OC model). The fact that the NLL+latent model becomes increasingly mis-calibrated as the complexity of the dataset increases reflects the fact that the distribution of errors becomes less normal as chemical complexity increases. The assumption of normally-distributed error is likely relatively valid for datasets with a few elements and similar molecular configurations such as MD17, while the error distribution of the OC20 dataset deviates significantly, likely due to the inclusion of 55 elements and adsorbate-catalyst interactions that involve metallic, covalent, and ionic bonding.

As for scalability, both the training and prediction costs of CP as a UQ method are relatively low compared to alternate approaches and the training and inference cost of the underlying neural network potential models such as SNN and GemNet-OC on a large dataset. Training the SNN model on 1M data points from OC20 takes $\sim120$ GPU-hours, and training GemNet-OC on the full 134M OC20 dataset takes $\sim11,000$ GPU-hours. In contrast, the training time for CP, including building the KDTree and performing quantile regression on the calibration set, is on the order of minutes. For prediction, the time required for CP to calculate the distance in 64 dimensions of latent space for the 1M training data is 106 ms/image for the SNN model, 23\% of the cost it takes to compute the fingerprints and $\sim$150 times the cost of a forward pass through the neural network. For GemNet-OC, the original latent vector dimension of 264 was reduced by performing PCA with a cumulative explained variance of 90\%, resulting in a latent vector dimension of 155, saving $\sim 40\%$ time based on linear scaling. As shown in Figure \ref{fig:scalability_latent_dimension}a), it would be possible to decrease the computational cost further by decreasing the dimension of the final latent layer of SNN models, or by utilizing PCA to further reduce the vector dimension for GemNet-OC. However, these strategies may negatively impact the accuracy of the model or the sharpness of the CP uncertainties, and this level of hyperparameter optimization is beyond the scope of this work. To provide a direct comparison with the SNN model and to reduce computational cost we chose to use the same dataset size for training CP+latent in the SNN and GemNet-OC models (1M randomly sampled points). The resulting time required for the CP prediction on GemNet-OC was 378 ms/image, which is $\sim$7 times the time needed for a forward pass of the GemNet-OC model\cite{GemNet-OC:Datasets}. Re-training a highly complex model such as GemNet-OC on a large 134M dataset is computationally expensive, making it impractical to utilize ensemble or dropout methods, and both ensemble or dropout will require multiple forward passes for prediction, making the inference cost comparable to or greater than CP+latent. This highlights the advantage of the CP method as a portable and scalable UQ method for pre-trained neural network models.


\section{Conclusion}

The CP method offers calibrated and sharp uncertainty estimates during prediction with relatively low computational costs compared to other heuristic methods. The latent space representation offers sharper uncertainty estimates compared to the feature space under CP. The CP+latent method can also be readily ported to trained neural network models, including deep graph-convolutional networks, showcased here for the GemNet-OC model. Additionally, the latent space representation allows a reduction in the computational costs of UQ by reducing the latent dimensions with a funnel-shaped neural network architecture. The CP+latent method yields higher uncertainty for out-of-distributional data compared to in-distributional ones as desired, suggesting the potential to be adopted in  active learning schemes to effectively explore the extrapolation region. However, the CP+latent method does have limitations since it may not be calibrated if the i.i.d assumption is violated due to out-of-distribution predictions or small training sample sizes, and the approach does not provide a direct way to estimate the uncertainty associated with specific atoms. Nonetheless, the CP+latent approach is a robust and scalable route for predicting uncertainty in MLFFs, and the general applicability of CP makes it a relevant calibration strategy for a range of other uncertainty heuristics, such as the standard deviations of the ensemble and dropout approaches or Bayesian models\cite{Angelopoulos2021}. We expect that the CP approach will find utility in active learning schemes for on-the-fly force field construction and materials discovery, particularly in cases where uncertainty estimates are needed for complex pre-trained models. 

\section{Acknowledgements}

The  authors  are  grateful  for funding from the U.S. Department of Energy’s Basic Energy Science, Computational Chemical Sciences Program Office, under Award No. DE-SC0019441. We also appreciate helpful discussions with Prof. Andrew A. Peterson.

\section{Data availability}

The data that support the findings of this study are openly available.

\clearpage
\newpage
\setcounter{figure}{0}
\renewcommand{\thefigure}{S\arabic{figure}}
\setcounter{table}{0}
\renewcommand{\thetable}{S\arabic{table}}

\section{Appendix: Code Example and Code Availability}

The listing below demonstrates the class,  \texttt{ConformalPrediction}, can be used by importing from Python package \texttt{AmpTorch} and fitting calibration data to predict the uncertainty.

\lstset{language=Python}
\lstset{frame=lines}
\lstset{caption={Conformal prediction implemented in \texttt{AmpTorch} to compute the prediction sets on a list of heuristic uncertainty metrics of test data.}}
\lstset{label={lst:code_cp}}
\begin{lstlisting}
# Step 1
# set up the conformal prediction model
model_cp = ConformalPrediction(calib_residual, calib_heuristic, alpha=alpha)
"""
class ConformalPrediction
Args:
    Input
    ``calib_residual``: 1D numpy.array. size: (num. of calibration data, ) 
    The absolute differences between the ground truths and predicted values, 
    | Y - Y_hat | 
    ``calib_heuristic``: 1D numpy.array. size: (num. of calibration data, ) 
    The scalar heuristics of calibration data, in this case, the computed 
    distances to training data. 
    ``alpha``: float, from 0 to 1. Default value set to 0.05 for 95% 
    confidence level. 
    ``test_uncertainty``: 1D numpy.array. size: (num. of test data,)
"""

# Step 2
# predict uncertainty given the calculated scalar uncertainty 
# heuristics of test data
test_uncertainty, qhat = model_cp.predict(test_heuristic)
#``test_heuristic``: 1D numpy.array. size: (num. of test data, )
#                    The uncertainty associated with each test data, 
#                    given the confidence level defined as (1-alpha)100%. 
#``qhat``: float. The (n+1)(1 - \alpha)/n quantile of score function, 
#                                                      |Y - Y_hat|/d(X). 

# Step 3
# return symmetric prediction sets as a tuple of lower and
# upper bound respectively. 
prediction_sets = (test_residual - test_uncertainty, 
                   test_residual + test_uncertainty) 
\end{lstlisting}

Here below, we list the resources for the implementation of conformal prediction with distances in the latent space.

\textbf{\texttt{AmpTorch} Python Module}

\begin{itemize}
    \item GitHub repository for installation and documentation: 
\url{https://github.com/ulissigroup/amptorch}
    \item Example script for a wrapper class, \texttt{ConformalPredictionLatentSpace}, with the implementation to fit and predict uncertainty on test set with a trained model: \url{https://github.com/ulissigroup/amptorch/blob/master/examples/GMP/GMP_uncertainty_example.py}
\end{itemize}

\newpage

\section{Appendix: Distance metrics}

\begin{table}[!htb]
    \centering
    \caption{Calibration and sharpness of three different distance metrics of the CP+latent method tested on the QM9 dataset. The NNFF model is trained on 50k data. $\alpha = 0.1$ and the expected confidence level is therefore 90\%. We implemented the Euclidean distances throughout the discussion following the convention by Janet et al\cite{PaulJanet2019}. }
    \label{tab:distance_metrics}
    \begin{tabular}{c c c}
    \hline
       Distance metric  &  Observed confidence level & $sha_{90\%}$ [eV/sys] \\
    \hline
       Euclidean ($p=2$) & 89.4\% & 0.0795  \\
       Manhattan ($p=1$) & 89.1\% & 0.0789 \\
       Chebyshev ($p=\infty$) & 89.6\% & 0.0802 \\
    \hline
    \end{tabular}
\end{table}

\section{Appendix: QM9}

\begin{table}[htb!]
    \centering
    \caption{Tabulated expected and observed confidence levels for different UQ methods on QM9 dataset. }
    \vspace{0.5em}
    \begin{tabular}{c | c c c | c c c }
        \hline
        \thead{Method} & \thead{Expected \\ conf. level} & \thead{Observed \\ conf. level} & \thead{Difference} & \thead{Expected \\ conf. level} & \thead{Observed \\ conf. level} & \thead{Difference} \\
        \hline \hline
        Ensemble    &   68\%    &   55\%    &   +13\% 
        &   95\%    &   83\%    &   +12\% \\
        Dropout     &   68\%    &   50\%    &   +18\% 
        &   95\%    &   77\%    &   +18\% \\
        NLL+feature &   68\%    &   90\%    &   -22\% 
        &   95\%    &   98\%    &   -3\% \\
        NLL+latent  &   68\%    &   78\%    &   -10\% 
        &   95\%    &   97\%    &   -2\% \\
        CP+feature  &   68\%    &   66\%    &   +2\% 
        &   95\%    &   95\%    &   0\% \\
        CP+latent   &   68\%    &   65\%    &   +3\% 
        &   95\%    &   95\%    &   0\% \\
         \hline
    \end{tabular}
    \label{tab:confidence_levels_qm9}
\end{table}

\begin{figure}[htb!]
	\centering
    \includegraphics[width=\linewidth]{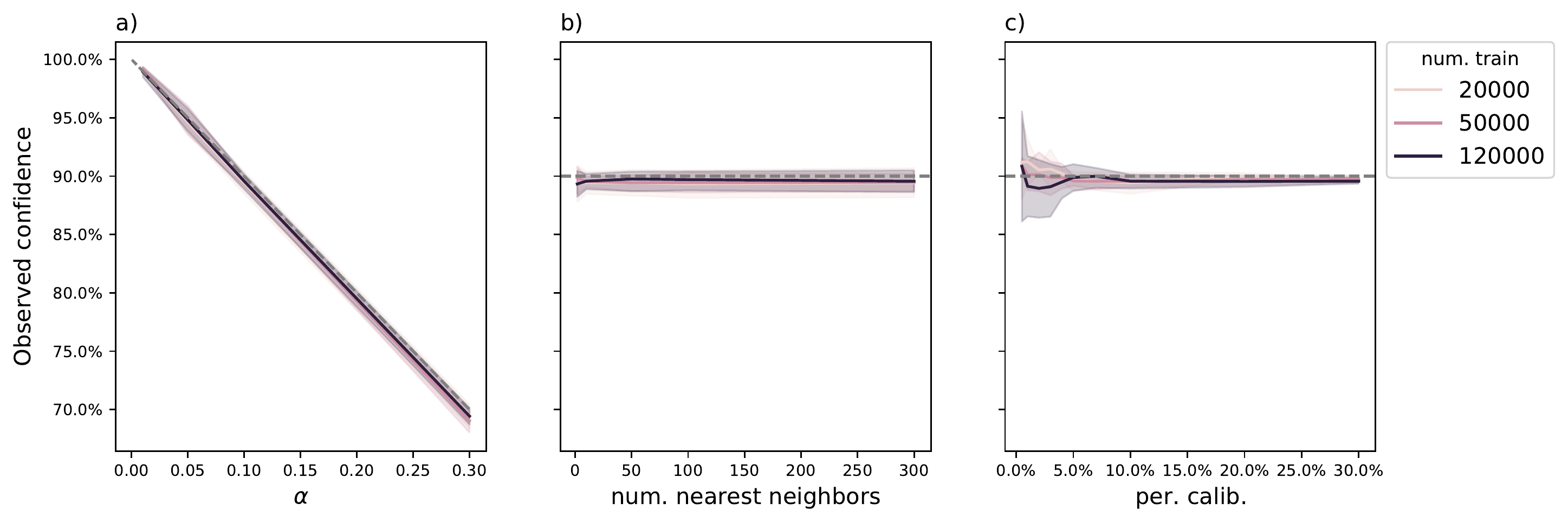}
    \caption{Effect of a) $\alpha$, b) num. of nearest neighbors, c) per. calib. on calibration with dataset QM9. The bands are one standard deviation calculated from 4 random seeds used to allocate calibration data. The expected confidence levels $(1-\alpha)100\%$ are plotted as gray dashed line for reference. The deviations with various $\alpha$ and num. of nearest neighbors are within $\sim$1\% given enough calibration data (per. calib. = 10\%). More deviation ($\sim$5\%) is observed with very few per. calib. (0.5\%, 1\%). After increasing per. calib. to $>10\%$, the deviation is $\sim$1\% consistently. The default values for $\alpha$ num. of nearest neighbors, and per. calib. are 0.1, 10, and 10\% respectively. 
    }
    \label{fig:calibration_parameter_qm9}
\end{figure}

\begin{figure}[hbt!]
	\centering
    \includegraphics[width=0.7\linewidth]{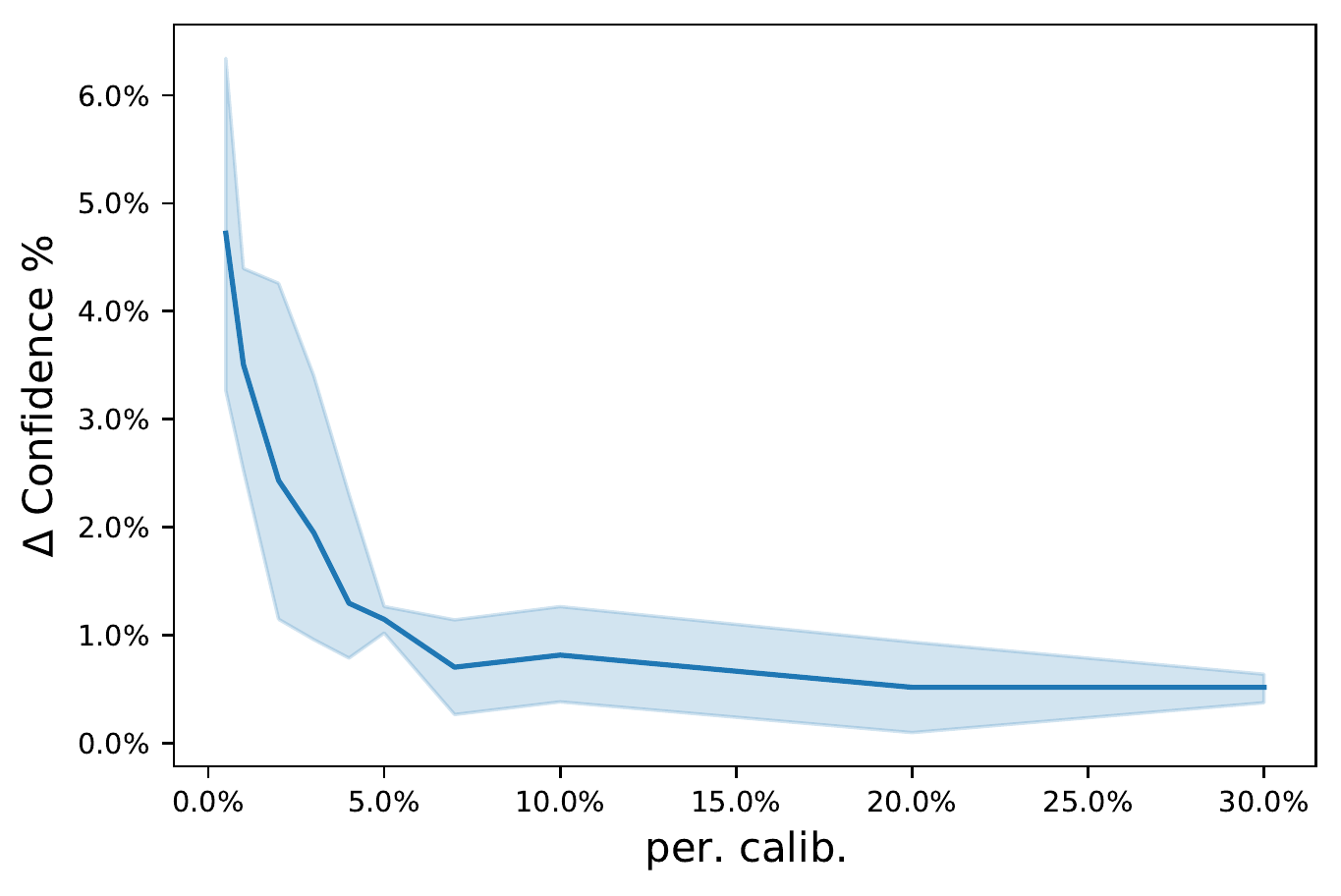}
    \caption{Deviation from expected confidence levels with respect to different per. calib. on QM9 with 120,000 trianing data. Num. of nearest neighbors is 10, and $\alpha$ is 0.1. The bands are generated by 4 different random seeds to uniformly select calibration data based on different numbers of per. calib. The deviation is $\sim$ 5\% at 0.5\% per. calib. and stays $\sim$ 1\% at per. calib. $>$10\%. }
    \label{fig:calibration_deviation_per_calib_qm9}
\end{figure}

In Figure \ref{fig:calibration_parameter_qm9}, we further break down the analysis and plot the observed confidence levels against model hyperparameters. The hyperparameter that affects calibration the most is per. calib. as shown in Figure \ref{fig:calibration_parameter_qm9}c). This is expected since the validity of underlying i.i.d. assumption depends on whether the amount of calibration data can reliably represent the test data. The deviation averaged over different selection of calibration data is $\sim$ 5\% with 0.5\% per. calib. (54 calibration images), and becomes $\sim$ 1\% when per. calib. is larger than 10\% (Figure \ref{fig:calibration_deviation_per_calib_qm9}). Given enough calibration data (per. calib. = 10\%), the observed confidence levels are calibrated to expected one within $\sim$ 2\% with various $\alpha$'s and numbers of nearest neighbors as shown in Figure \ref{fig:calibration_parameter_qm9}a) and b). These results confirm the robustness of the calibration of the CP method under a wide range of scenarios given enough calibration data (rule of thumb: per. calib. $>$ 10\%).

\begin{figure}[htb!]
	\centering
    \includegraphics[width=\linewidth]{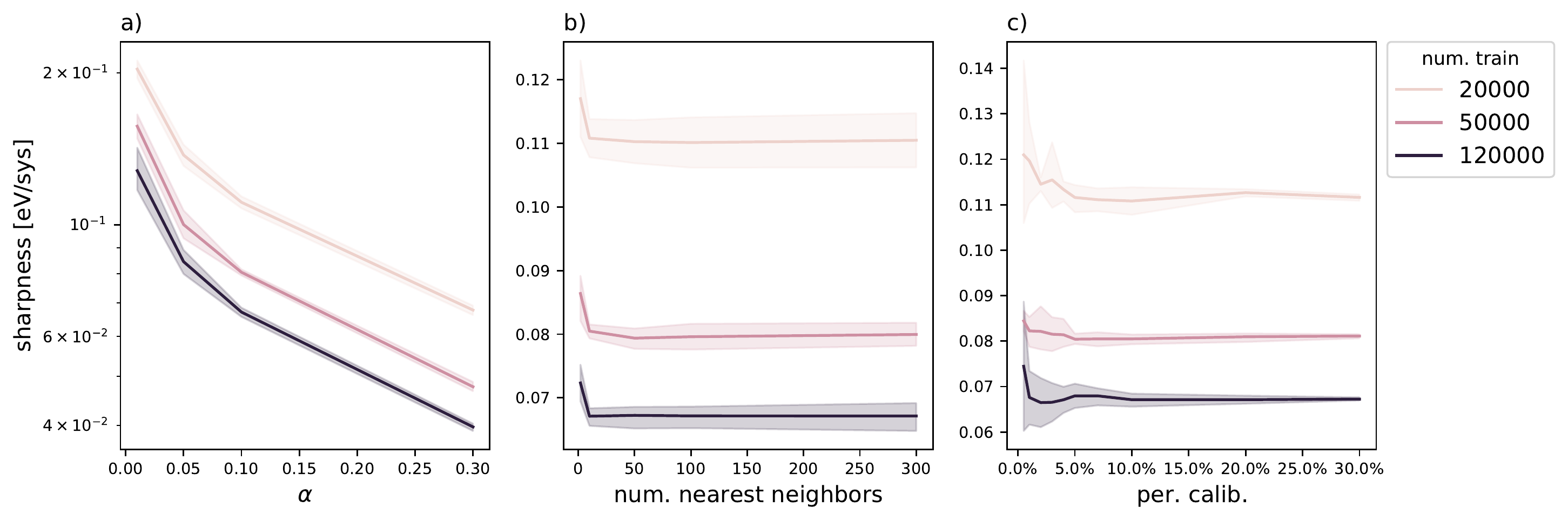}
    \caption{Effect of a) $\alpha$, b) num. of nearest neighbors, c) per. calib. on sharpness with dataset QM9. The bands are one standard deviation calculated from 4 random seeds used to allocate calibration data. Models with higher accuracy (models trained with more training data) tend to be sharper than models with lower accuracy. As $\alpha$ increases, the confidence level decreases, therefore better sharpness. Num. of nearest neighbors, at least 10, to ensure better sharpness. The default values for $\alpha$, num. of nearest neighbors, and per. calib. are 0.1, 10, and 10\% respectively. }
    \label{fig:sharpness_parameter_qm9}
\end{figure}

We then move on to characterize how the model hyperparameters affect the sharpness of the UQ method in Figure \ref{fig:sharpness_parameter_qm9}. First, the model accuracy determines the overall level of sharpness, as the more number of training data, the better sharpness there is for all three model hyperparameters. Second, $\alpha$ has a substantial effect on sharpness. As $\alpha$ changes from 0.001 to 0.3, the sharpness changes by an order of magnitude. This is expected, since increasing the confidence level will necessarily decrease the sharpness. Third, a threshold number of nearest neighbors is required to obtain sharp estimates of uncertainty under the same model accuracy (Figure \ref{fig:sharpness_parameter_qm9}b), and the threshold number is $\sim$ 10, consistent with previous findings. Adding more nearest neighbors is not helpful after the threshold. Lastly, the percent of calibration data allocated affects only the variation in sharpness due to randomness, especially when the amount of calibration data is small, with the sharpness converging at roughly 10\% calibration data (consistent with the calibration results in Fig. \ref{fig:calibration_parameter_qm9}). 

\begin{figure}[htb!]
	\centering
    \includegraphics[width=\linewidth]{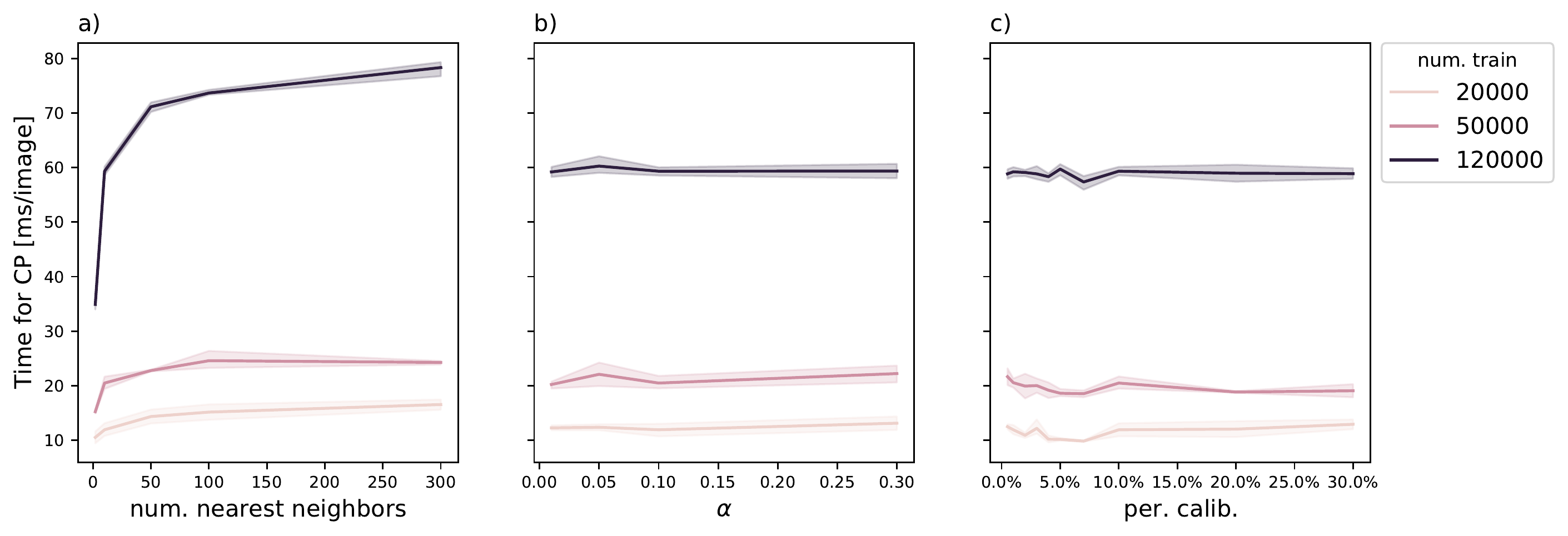}
    \caption{Effect of a) $\alpha$, b) num. of nearest neighbors, c) per. calib. on sharpness with dataset QM9. The bands are one standard deviation calculated from 4 random seeds used to allocate calibration data. The default values for $\alpha$, num. of nearest neighbors, and per. calib. are 0.1, 10, and 10\% respectively.}
    \label{fig:cp_times_parameter_qm9}
\end{figure}

Another critical factor is how the model hyperparameters affect the overall scalability or run-time for CP models. The dominant factor of scalability is the number of nearest neighbors in the distance calculation because the time complexity of the KDTree algorithm is $O(D N \log N)$. Keeping the number of nearest neighbors low means less run-time. $\alpha$ is processed during the quantile regression, which takes up a negligible amount of time, and therefore does not affect the overall scalability (Figure \ref{fig:cp_times_parameter_qm9}b). The percent of calibration does not affect scalability because the data point is either labeled as calibration or test, following similar processes.

\clearpage
\newpage
\section{Appendix: MD17-Aspirin}

\begin{figure}[htb!]
	\centering
        \includegraphics{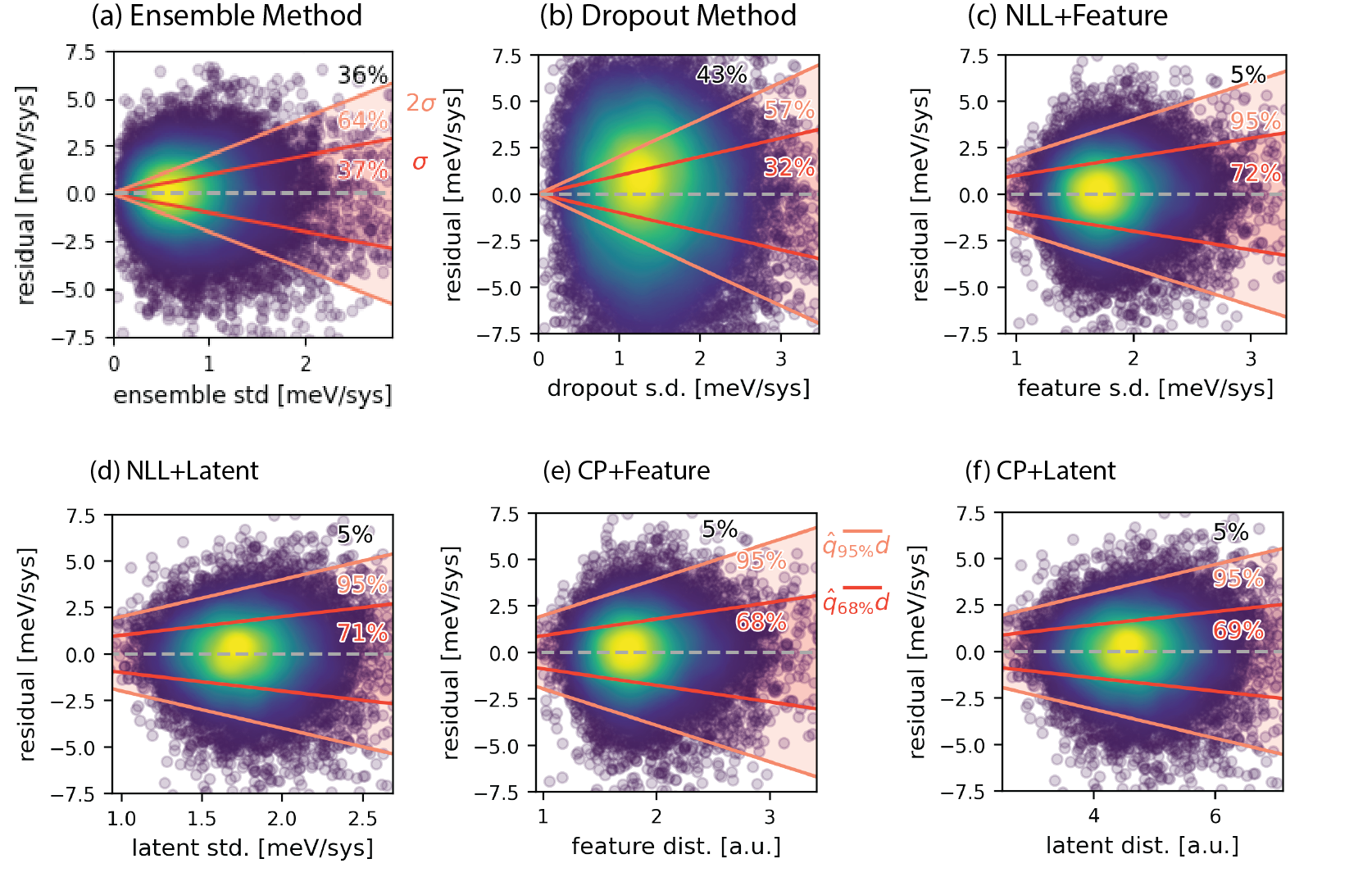}
    \caption{Method comparison on MD17 dataset. Residuals are the difference between ground truths and predictions. The color of the dots indicates the density of dots in close proximity from purple (low density) to yellow (high density) by KDE analyses. The expected values for printed three percent numbers in every plot are 68\% (red), 95\% (pink), and 5\% (black).  (a,b,c,d) the red line indicates the spread of one standard deviation coverage, and the pink line indicates two standard deviations. Under Gaussian distributional assumption, the sample coverage should be 68\%, 95\%, while the tested coverage for these four methods is off. (e,f) the red line indicates the prediction bandwidths by $1-\alpha=68\%$, and the pink line indicates the prediction bandwidths by $1-\alpha=95\%$. The tested coverage is close to the pre-defined expected confidence levels for both CP methods. }
    \label{fig:method_comparison_md17}
\end{figure}

\begin{figure}[htb!]
	\centering
    \begin{subfigure}[t]{0.3\textwidth}
		\centering
        \subcaption[short for lof]{Ensemble method}
        \includegraphics[width=\linewidth]{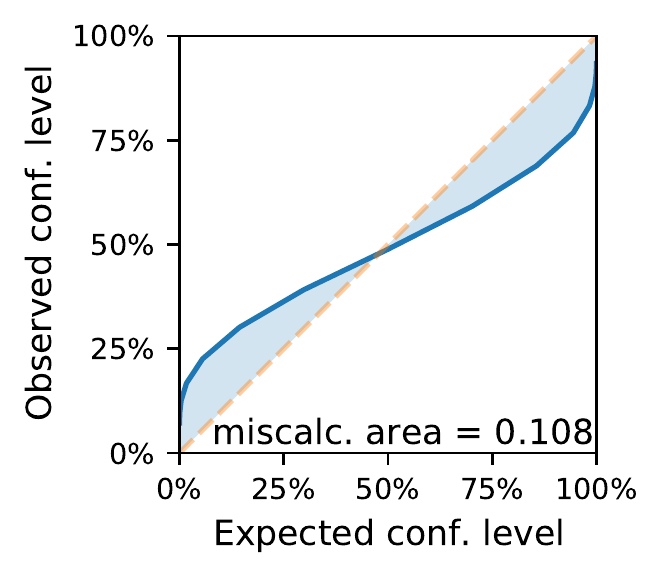}
	\end{subfigure}\hspace{0.005\textwidth}
    \begin{subfigure}[t]{0.3\textwidth}
		\centering
        \subcaption[short for lof]{Dropout method}
        \includegraphics[width=\linewidth]{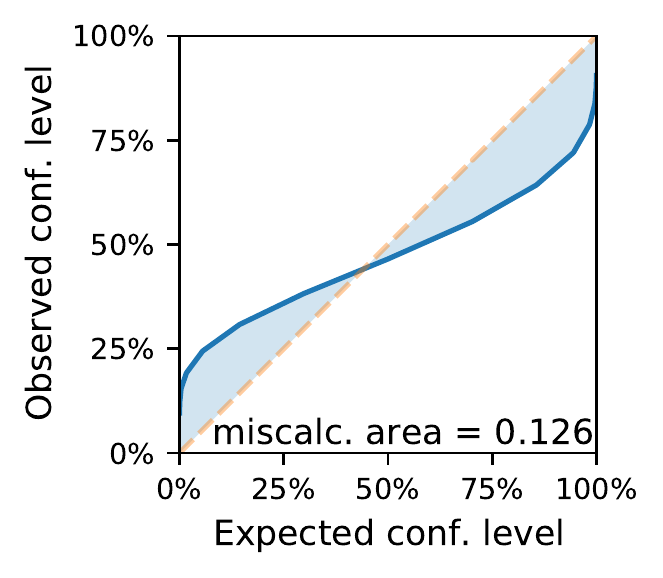}
	\end{subfigure}\hspace{0.005\textwidth}
    \begin{subfigure}[t]{0.3\textwidth}
		\centering
		\subcaption[short for lof]{NLL+feature}
        \includegraphics[width=\linewidth]{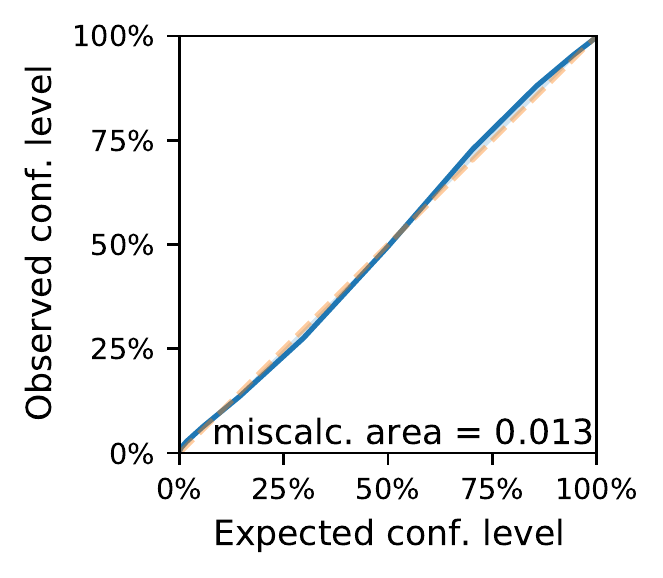}
	\end{subfigure}\hspace{0.005\textwidth}%
    \begin{subfigure}[t]{0.3\textwidth}
		\centering
		\subcaption[short for lof]{NLL+latent}
        \includegraphics[width=\linewidth]{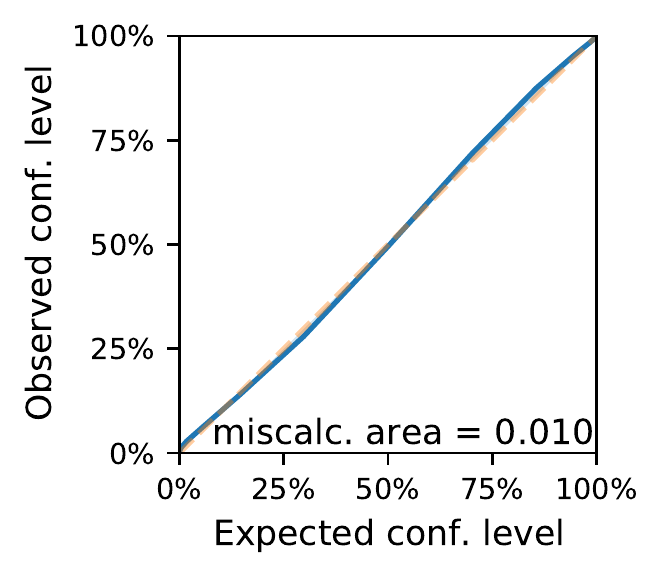}
	\end{subfigure}\hspace{0.005\textwidth}%
    \begin{subfigure}[t]{0.3\textwidth}
		\centering
		\subcaption[short for lof]{CP+feature}
        \includegraphics[width=\linewidth]{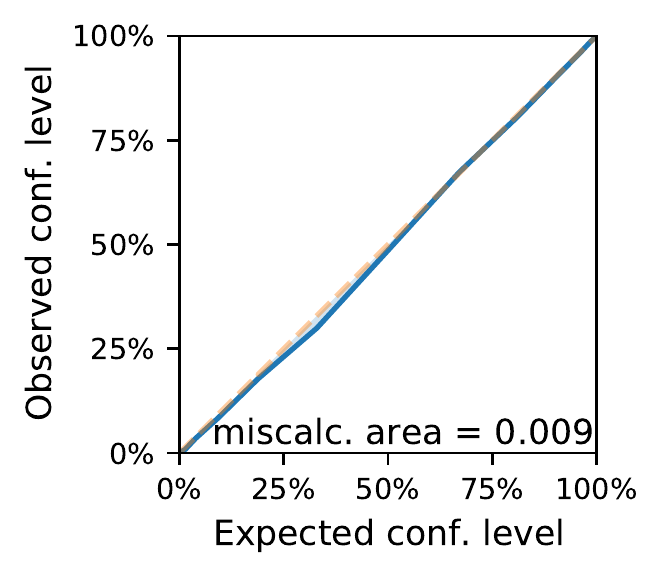}
	\end{subfigure}\hspace{0.005\textwidth}%
    \begin{subfigure}[t]{0.3\textwidth}
		\centering
		\subcaption[short for lof]{CP+latent}
        \includegraphics[width=\linewidth]{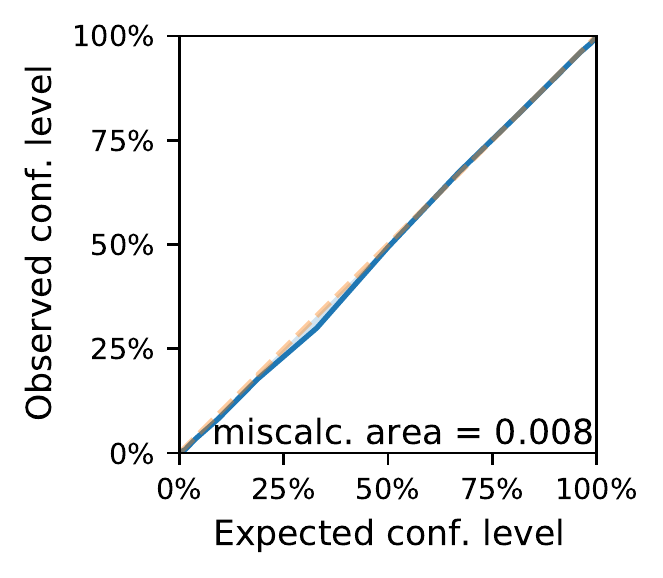}
	\end{subfigure}\hspace{0.005\textwidth}%
    \caption{Calibration curve comparison on MD17-Aspirin dataset. The orange dashed line is the ideal case where the observed confidence level always matches the expected confidence level. The solid blue line is the actual observation. The area between the solid blue line and the orange dashed line indicates the frequency the model is miscalibrated. At the bottom of every plot prints the values for miscalibrated areas. }
    \label{fig:calibration_curve_md17}
\end{figure}

\begin{figure}[htb!]
	\centering
    \includegraphics[width=0.9\linewidth]{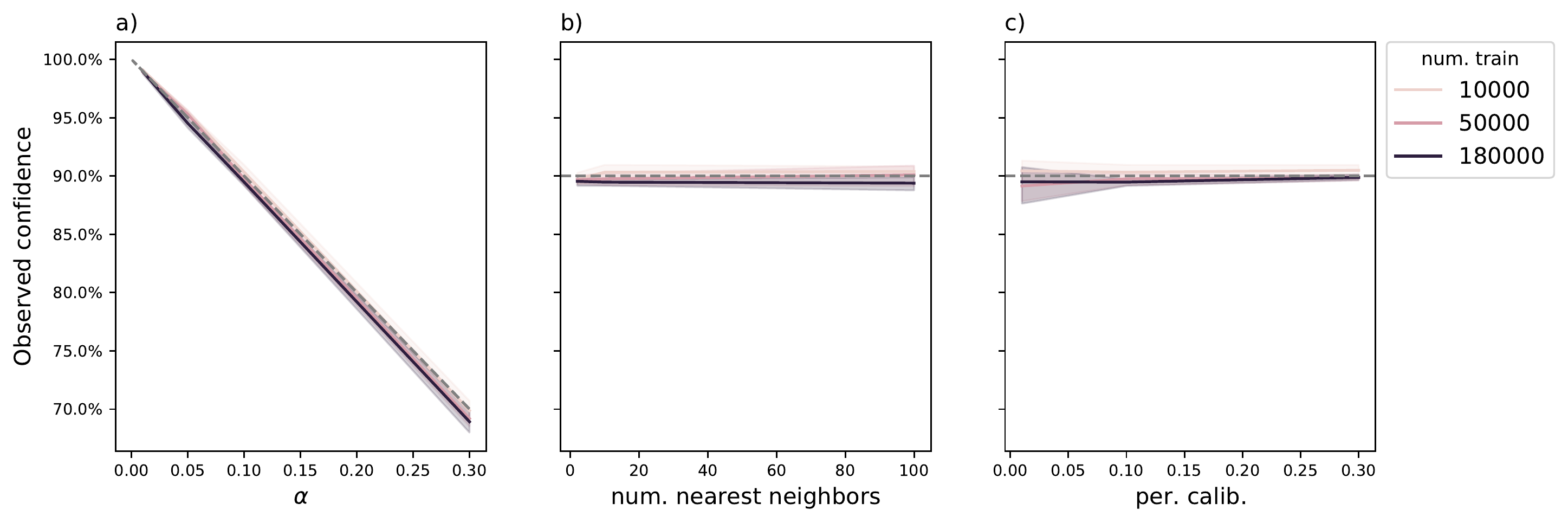}
    \caption{Effect of a) $\alpha$, b) num. of nearest neighbors, c) per. calib. on calibration with dataset MD17-Aspirin. The bands are one standard deviation calculated from 4 random seeds used to allocate calibration data. The default values for $\alpha$, num. of nearest neighbors, and per. calib. are 0.1, 10, and 10\% respectively.}
    \label{fig:calibration_parameter_md17}
\end{figure}

\begin{figure}[!t]
	\centering
    \includegraphics[width=0.9\linewidth]{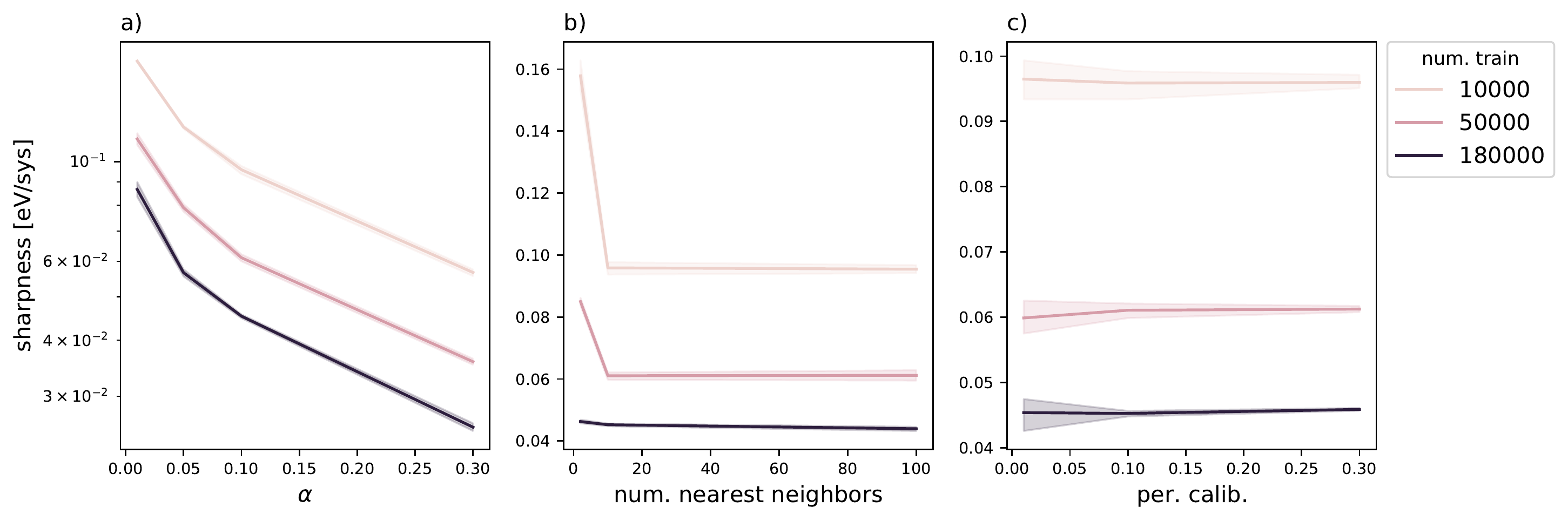}
    \caption{Effect of a) $\alpha$, b) num. of nearest neighbors, c) per. calib. on sharpness with dataset MD17-Aspirin. The bands are one standard deviation calculated from 4 random seeds used to allocate calibration data. The default values for $\alpha$, num. of nearest neighbors, and per. calib. are 0.1, 10, and 10\% respectively.}
    \label{fig:sharpness_parameter_md17}
\end{figure}

\clearpage
\newpage
\section{Appendix: OOD analysis on QM9}

\begin{figure}[htb!]
	\centering
    \includegraphics[width=
    \linewidth]{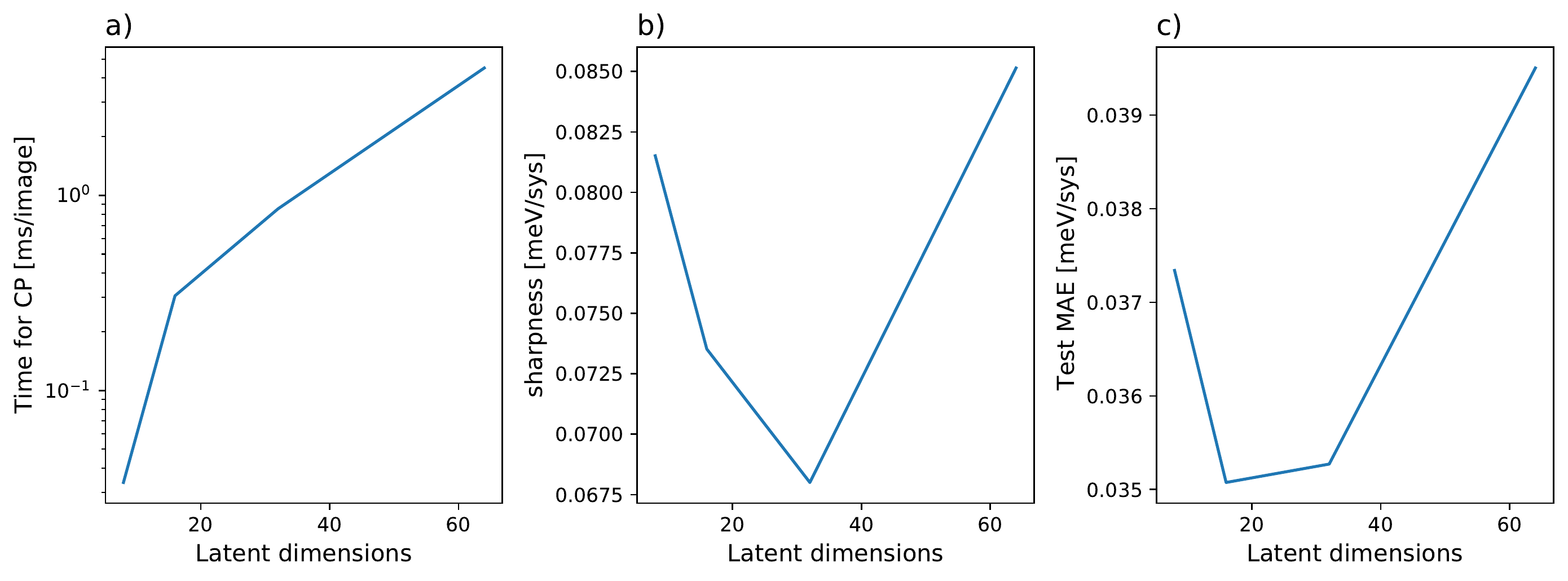}
    \caption{Scalability analysis when applying the CP+latent method to different number of latent dimensions on the QM9 dataset. a) The time it takes to perform CP+latent distance analysis on different neural network architectures with different dimension for last-layer latent spaces. b) Sharpness with different dimensions of last-layer latent spaces. c) The model accuracy in MAE with respect to different dimensions of last-layer latent spaces.}
    \label{fig:scalability_latent_dimension}
\end{figure}

\clearpage
\newpage

\section{Appendix: Iterative training on OOD F+ molecules of QM9}
\label{sec:ood_active_learning}

\begin{figure}[!htb]
    \centering
    \includegraphics[width=.9\textwidth]{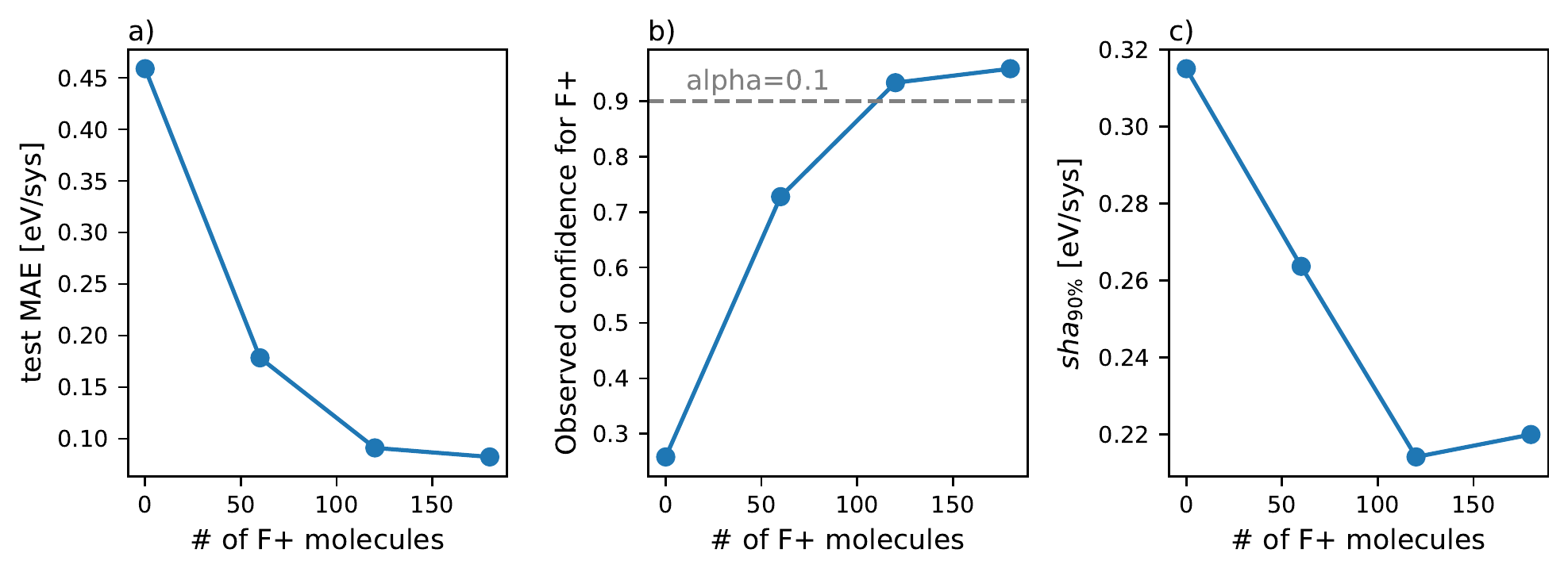}
    \caption{Iteratively training with the most uncertain molecules pooled from $\sim$ 2k F+ molecules of QM9. With an initial model trained only on F- molecules, we performed the CP+latent method with a calibration set with F- molecules and computed uncertainties on test F+ molecules. We added the top 60 uncertain F+ molecules (x-axis) to training data during every iteration and trained the corresponding model. a) MAE of the model trained with 50k F- molecules and the added F+ molecules. b) Observed confidence on test F+ molecules at $\alpha = 0.1$. The expected confidence is 0.9 or 90\%. c) Sharpness at confidence 90\% of the test F+ molecules. }
    \label{fig:ood_active_learning}
\end{figure}

\begin{figure}[htb!]
	\centering
    \begin{subfigure}[t]{0.32\textwidth}
		\centering
		\subcaption[short for lof]{\# of F+ = 0}
        \includegraphics[width=\linewidth]{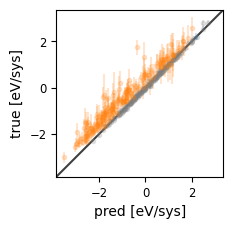}
	\end{subfigure}
    \begin{subfigure}[t]{0.32\textwidth}
		\centering
		\subcaption[short for lof]{\# of F+ = 60}
        \includegraphics[width=\linewidth]{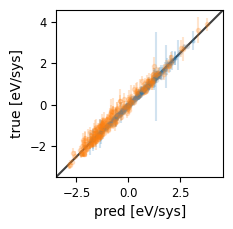}
	\end{subfigure}
    \begin{subfigure}[t]{0.32\textwidth}
		\centering
		\subcaption[short for lof]{\# of F+ = 120}
        \includegraphics[width=\linewidth]{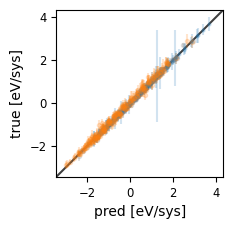}
	\end{subfigure}
    \caption{Parity plots for models trained with F- molecules and different numbers of F+ selected by uncertainty. DFT-calculated energies are reported as true energies in the y-axis, and the NN force field predicted energies as pred energies in the x-axis. Plots a), b), and c) correspond to the first, second, and third dots (models) in Figure \ref{fig:ood_active_learning}. Training F- molecules are displayed as gray dots with 90\% prediction bands. Among 110K F- molecules, 200 are randomly selected and plotted to aid in visualization. Training F+ molecules are all displayed as blue dots with 90\% prediction bands. Test F+ molecules are displayed as orange dots with 90\% prediction bands. Among $\sim$2K test F+ molecules, 200 are randomly selected and plotted to aid in visualization.}
    \label{fig:active_learning_parity}
\end{figure}

\clearpage
\newpage
\section{Appendix: OC20}

\begin{figure}[htb!]
	\centering
    \begin{subfigure}[t]{0.45\textwidth}
		\centering
		\subcaption[short for lof]{NLL+latent in SingleNN}
        \includegraphics[width=\linewidth]{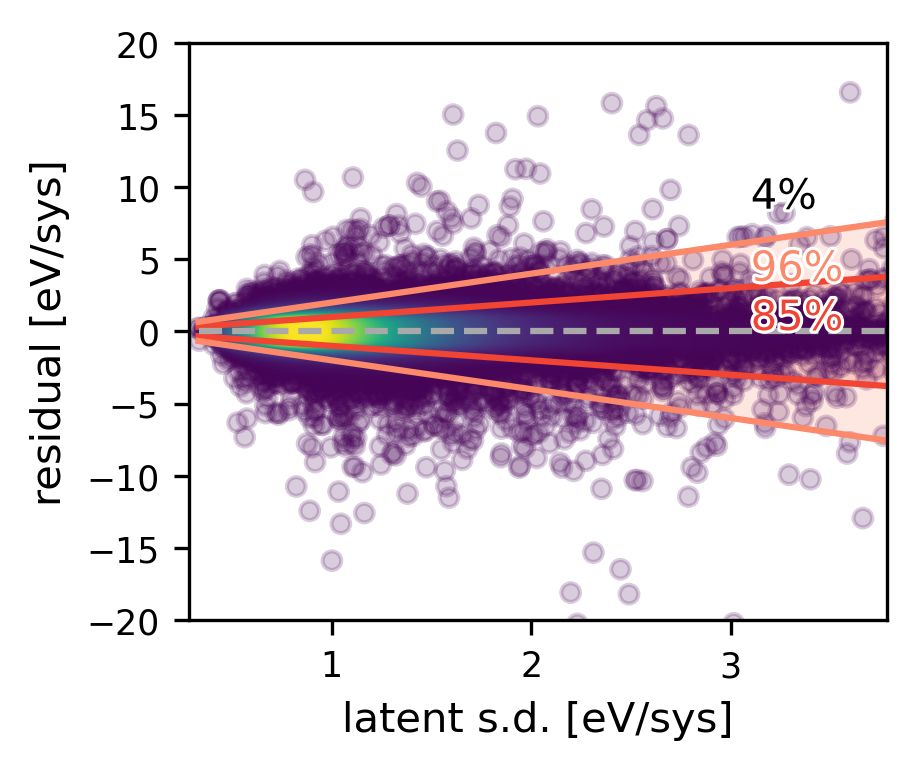}
        \label{fig:singlenn_nll_latent_mc_oc20}
	\end{subfigure}\hspace{0.005\textwidth}%
    \begin{subfigure}[t]{0.45\textwidth}
		\centering
		\subcaption[short for lof]{CP+latent in SingleNN}
        \includegraphics[width=\linewidth]{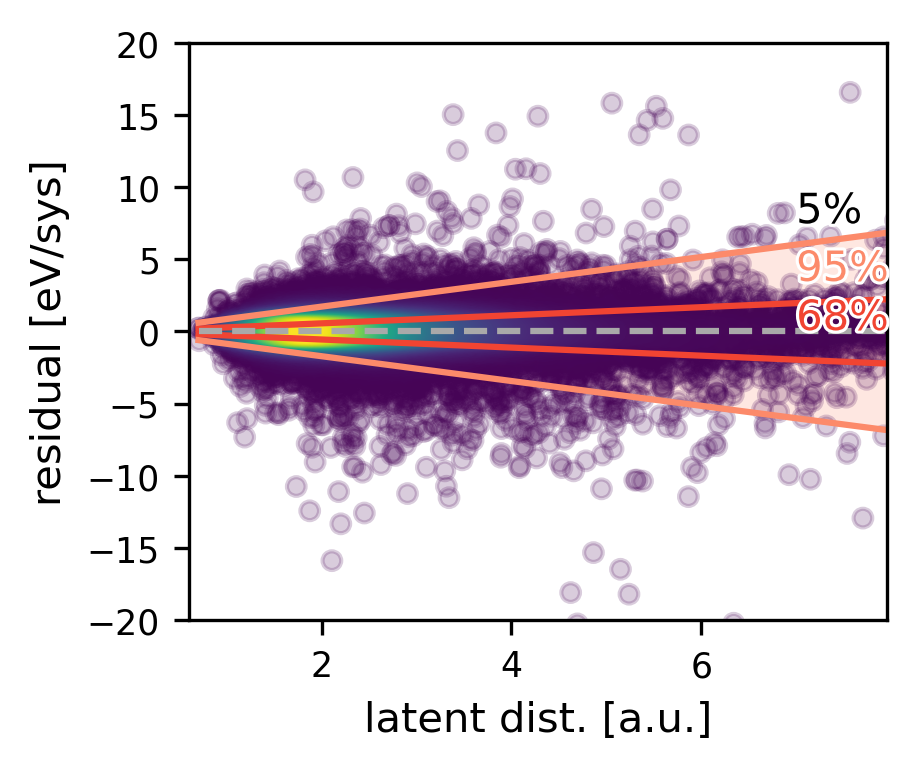}
        \label{fig:singlenn_cp_latent_mc_oc20}
	\end{subfigure}\hspace{0.005\textwidth}%
    \begin{subfigure}[t]{0.45\textwidth}
		\centering
		\subcaption[short for lof]{NLL+latent in GemNet-OC}
        \includegraphics[width=\linewidth]{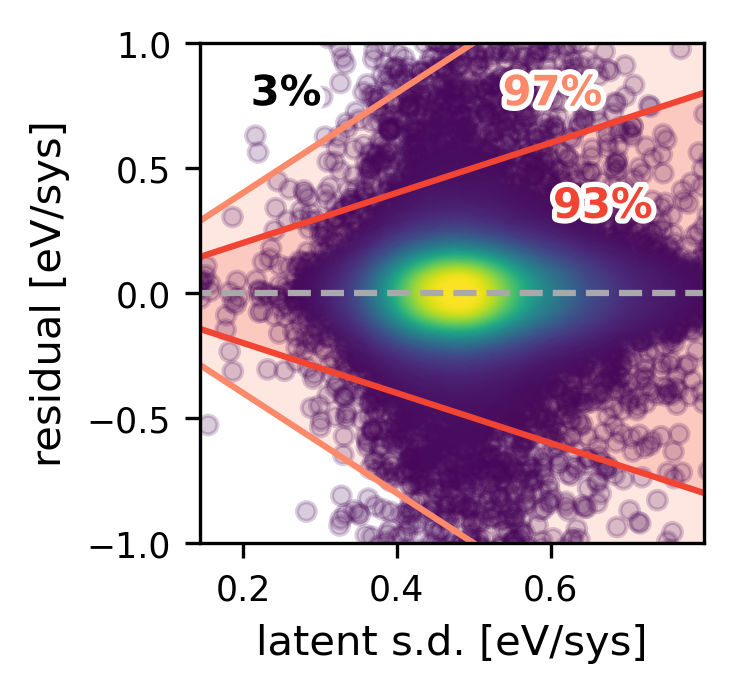}
        \label{fig:gemnet_nll_latent_mc_oc20}
	\end{subfigure}\hspace{0.005\textwidth}%
    \begin{subfigure}[t]{0.45\textwidth}
		\centering
		\subcaption[short for lof]{CP+latent in GemNet-OC}
        \includegraphics[width=\linewidth]{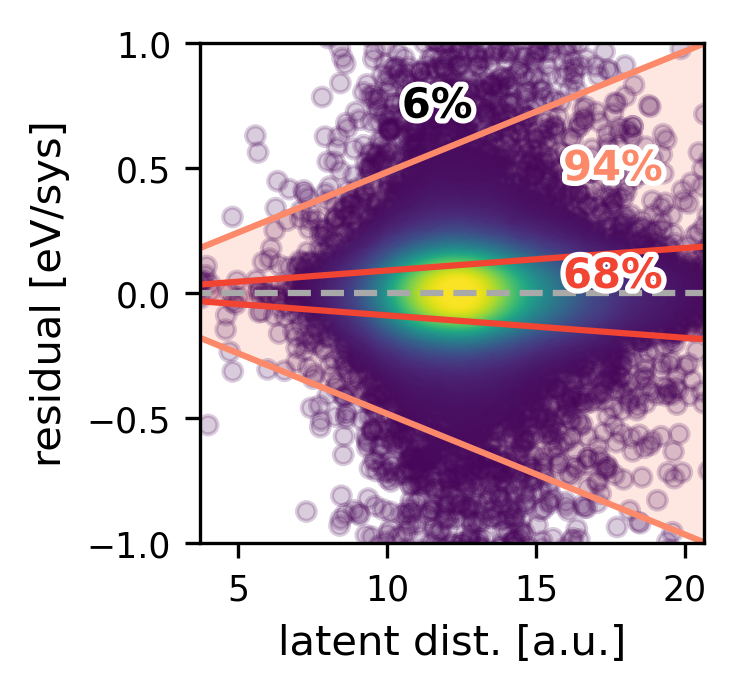}
        \label{fig:gemnet_cp_latent_mc_oc20}
	\end{subfigure}\hspace{0.005\textwidth}%
    \caption{Comparison of NLL+latent distances and CP+latent distances on the 1-million OC20 dataset. 5\% randomly selected data are visualized here, but the observed confidence levels are calculated for all test data. (a) Uncertainty estimated by NLL+latent method as the standard deviations. The region bounded between the red lines is within one standard deviation, and the region bounded between the orange line is within two standard deviations. (b) Uncertainty estimated by CP+latent method as the prediction sets at 68\% and 95\% confidence levels. The region bounded between red lines is 68\% expected confidence level and the oranges lines 95\%. The observed confidence levels match the expectation. }
    \label{fig:method_comparison_oc20}
\end{figure}

\clearpage
\newpage
\bibliographystyle{unsrt}  

\end{document}